\documentclass[%
12pt,
aip,
cha,
amsmath,amsfonts,amssymb,
author-numerical,
reprint,
floatfix
]{revtex4-2}

\usepackage[utf8]{inputenc}
\usepackage[T1]{fontenc}
\usepackage{graphicx}
\usepackage{bm}
\usepackage{mathrsfs}
\usepackage{xcolor}

\def\pct{\%}
\def\bpi{\mathbf p}
\renewcommand{\d}[1]{\ensuremath{\operatorname{d}\!{#1}}}
\DeclareMathOperator*{\m}{area}
\DeclareMathOperator{\spn}{span}
\DeclareMathOperator{\Id}{Id}%

\begin{document}

\title{Transition paths of marine debris and the stability of the garbage patches}

\author{P.\ Miron} \email{pmiron@miami.edu} \affiliation{Department
of Atmospheric Sciences, Rosenstiel School of Marine and Atmospheric
Science, University of Miami, Miami, Florida, USA}

\author{F.\ J.\ Beron-Vera} \affiliation{Department of Atmospheric
Sciences, Rosenstiel School of Marine and Atmospheric Science,
University of Miami, Miami, Florida, USA}

\author{L.\ Helfmann} \affiliation{Institute of Mathematics, Freie
Universit\"at Berlin, Berlin, Germany} \affiliation{Department of
Modeling and Simulation of Complex Processes, Zuse-Institute Berlin,
Berlin, Germany} \affiliation{Complexity Science Department, Potsdam
Institute for Climate Impact Research, Potsdam, Germany}

\author{P.\ Koltai} \affiliation{Institute of Mathematics, Freie
Universit\"at Berlin, Berlin, Germany}

\date{\today}

\begin{abstract}
 We used transition path theory (TPT) to infer ``reactive'' pathways
 of floating marine debris trajectories. The TPT analysis was applied
 on a pollution-aware time-homogeneous Markov chain model constructed
 from trajectories produced by satellite-tracked undrogued buoys
 from the NOAA Global Drifter Program. The latter involved coping
 with the openness of the system in physical space, which further
 required an adaptation of the standard TPT setting.  Directly
 connecting pollution sources along coastlines with garbage patches
 of varied strengths, the unveiled reactive pollution routes represent
 alternative targets for ocean cleanup efforts. Among our specific
 findings we highlight: constraining a highly probable pollution
 source for the Great Pacific Garbage Patch; characterizing the
 weakness of the Indian Ocean gyre as a trap for plastic waste; and
 unveiling a tendency of the subtropical gyres to export garbage
 toward the coastlines rather than to other gyres in the event of
 anomalously intense winds.
\end{abstract}

\pacs{02.50.Ga; 47.27.De; 92.10.Fj}

\maketitle

\textbf{
 Given a Markov chain, namely, a model describing the stochastic
 state transitions in which the transition probability of each state
 depends only on the state attained in the previous event, transition
 path theory (TPT) provides a rigorous approach to study the
 statistics of transitions from a set of states to another, possibly
 disconnected set of states. Envisioning the motion of floating
 debris as described by a Markov chain that accounts for the ability
 of coastal states to ``pollute the oceans,'' TPT is employed to
 unveil ``reactive'' pathways representing direct transitions from
 potential release locations along the shorelines to accumulation
 sites across the world ocean. These include the subtropical gyres,
 whose strength in this context is investigated.
}

\section{Introduction}

The long-term fate of satellite-tracked drifting buoys from the
NOAA Global Drifter Program \cite{Lumpkin-Pazos-07} is characterized
by a tendency to form clusters in the oceans' subtropical gyres
\cite{vanSebille-etal-12, Maximenko-etal-12} that resemble great
garbage patches.\cite{Cozar-etal-14} The development of such clusters,
most evidently in the case of undrogued (i.e., without a sea anchor)
drifters,\cite{Beron-etal-16} has been explained \cite{Beron-etal-16,
Beron-etal-19-PoF, Beron-20} as the result of the combined action
on the drifters of converging ocean currents and winds mediated by
their inertia, which prevent them from adapting their velocities
to that of the carrying water--air flow system.

The tendency of the drifters to cluster in the long run enables a
probabilistic description of their dynamics using results from
ergodic theory \cite{Lasota-Mackey-94} and Markov chains,
\cite{Bremaud-99, Norris-98} which form the basis for approximating
asymptotically invariant sets using so-called \emph{set-oriented
methods}.\cite{Dellnitz-Hohmann-97, Dellnitz-Junge-99,
Froyland-Dellnitz-03, Koltai-10} This approach places the focus on
the evolution of probability densities, which, unlike individual
trajectories, represent robust features of the dynamics. Central
to this measure-theoretic characterization is the transfer operator
and the transition matrix, its discrete version resulting by covering
the phase space with boxes, which represent the states of the
associated Markov chain.

Such a probabilistic description has been applied on \emph{simulated}
drifter trajectories,\cite{Froyland-etal-14} suggesting a
characterization of great garbage patches as almost-invariant
attracting sets with corresponding basins of attraction spanning
areas as large as those of the geographic ocean basins. While the
latter suggests a strong influence of the regions collecting marine
debris on their global transport, it does not provide information
on pollution routes.

The goal of this paper is to unveil such routes from \emph{observed}
drifter trajectories. This is done by applying \emph{transition
path theory} (\emph{TPT}).\cite{VandenEijnden-06, Metzner-etal-06,
Weinan-VandenEijnden-06, Metzner-etal-09, Weinan-VandenEijnden-10}
Developed to investigate transition pathways in complex nonlinear
stochastic systems, TPT provides a statistical characterization of
the ensemble of ``reactive'' trajectories, namely, pieces of
trajectories along which \emph{direct} transitions between two sets
$A$ and $B$ in phase space take place. The TPT terminology is
borrowed from statistical mechanics and physical chemistry, for
which TPT was originally developed to study chemical reactions from
reactants $A$ to products $B$, as an improvement for earlier
approaches such as transition state theory \cite{Wigner-38} and
transition path sampling.\cite{Pratt-86} Since then, the TPT framework
has also been applied to studying molecular conformation changes
\cite{Noe-etal-09, Voelz-etal-10} and transitions in climate
models.\cite{Lucente-etal-19, Finkel-etal-20} We here present, to
the best of our knowledge, the first oceanographic application.

By constructing a Markov chain for debris motion and then identifying
coastline boxes in the ocean covering with reactant states $A$, and
boxes in several ocean locations including the subtropical gyres
with product states $B$, we use TPT to infer pollution pathways in
the global ocean.  The Markov chain model accounts for the ability
of coastal boxes (states) to ``pollute the oceans.'' This involves
adding an artificial state to the chain where all outflow goes in
and all inflow comes from (in an manner that differs from prior
approaches\cite{Froyland-etal-14c, Lunsmann-etal-18}).  By setting
$A$ to a single garbage patch and $B$ as the union of the other
garbage patches, we can also assess the strength of the patches.

The rest of the paper is organized as follows. The ergodic-theory
setup for closed systems is presented in Sec.\ \ref{sec:setup}.  An
adaptation of the theory for open systems in discussed in
Sec.\ \ref{sec:closure}. The main results of TPT are reviewed in
Sec.\ \ref{sec:tpt}, both for closed systems (Sec.\ \ref{ssec:tpt_closed})
and an extension for open domains  (Sec.\ \ref{ssec:tpt_open}).  The
Markov-chain model for ocean pollution is constructed in
Sec.\ \ref{sec:MCM} from satellite-tracked drifter trajectories.
This entails coping with a number of issues, previously not
encountered, partially addressed, or overlooked,\cite{Miron-etal-17,
Miron-etal-19-Chaos, Miron-etal-19-JPO, Olascoaga-etal-18,
Beron-etal-20-Chaos} these include: zonal connectivity; spurious
communication between ocean basins; and nonobserved communication;
as well as incorporating pollution sources near the coast.  In Sec.\
\ref{sec:asym} time-asymptotic aspects of the chain dynamics are
investigated, suggesting prospects for garbage patches yet to be
directly observed.  The TPT analysis is applied in Sec.\
\ref{sec:debris_paths}.  This reveals pollution routes into the
garbage patches, which represent alternative targets for ocean
cleanup efforts.\cite{Morrison-etal-19} Finally, a summary and the
conclusions of the paper are presented in Sec.\ \ref{sec:conclusion}.

\section{Setup for closed dynamical systems}\label{sec:setup}

Let us assume that floating debris trajectories are described by a
time-homogeneous stochastic process in continuous space $\mathcal
X\subset \mathbb{R}^2$ and observed at discrete times $nT$,
$n\in\mathbb Z$. Its transition probabilities are controlled by a
stochastic kernel $K(x,y) \ge 0$ such that $\smash{\int_{\mathcal
X} K(x,y)\d{y} = 1}$ for all $x$ in phase space $\mathcal X$,
representing the world ocean basin. The stochastic
kernel is time-independent since the time-homogeneity of the process
implies that the rules governing the process at any time are the
same. It is convenient to think of $\mathcal X$ as a measure space,
i.e., a set equipped with a $\sigma$-algebra of subsets measured
by (normalized) area. Then a probability density $f(x) \ge 0 $,
$\smash{\int_{\mathcal X} f(x) \d{x} = 1}$, describing the distribution
of the random position $X_{nT}$ at any time $nT$ evolves to the
distribution
\begin{equation}
 \mathscr Pf(y) := \int_{\mathcal X} K(x,y)f(x)\d{x}
\end{equation}
at time $(n+1)T$, which defines a Markov operator $\mathscr P :
L^1(\mathcal X) \circlearrowleft$ generally known as a \emph{transfer
operator}.\cite{Lasota-Mackey-94}

To infer the action of $\mathscr P$ on a discretized space one can
use a Galerkin projection referred to as Ulam's method.\cite{Ulam-60,
Kovacs-Tel-89, Koltai-10} This consists of covering the phase space
$\mathcal X$ with $N$ connected boxes $\smash{\{B_i\}_{i\in S}}$,
$S := \{1,\dotsc,N\}\subset \mathbb Z^+$, disjoint up to zero-measure
intersections, and projecting functions in $L^1(\mathcal X)$ onto
the finite-dimensional space spanned by indicator functions on the
boxes $V_N := \spn\smash{\big\{\frac{\mathbf
1_{B_i}(x)}{\m(B_i)}\big\}_{i\in S}}$ where $\mathbf 1_A(x) = 1$
if $x\in A$ and 0 otherwise. The discrete action of $\mathscr P$
on $V_N$ is described by a matrix $P = \smash{(P_{ij})_{i,j\in
S}}\in \smash{\mathbb{R}^{N\times N}}$ called a \emph{transition
matrix}. The transition matrix results from the projection
\cite{Miron-etal-19-JPO, Miron-etal-19-Chaos}
\begin{align}
 P_{ij} &:= \Pr(X_{(n+1)T}\in B_j \mid X_{nT} \in B_i)\nonumber\\
 &= \frac{1}{\m(B_i)}\int_{B_i}
 \int_{\!{B_j}} K(x,y) \d{x}\d{y} 
 \label{eq:Pdef}
\end{align}
and describes the proportion of probability mass in $B_i$  that
flows to $B_j$ during $T$. If one is provided with a large set of
observations $x_0$ and $x_T$ of $X_0$ and $X_T$, respectively, then
\eqref{eq:Pdef} can be estimated via counting the transitions in
the observed data, viz.,
\begin{equation}
 P_{ij} = \frac{C_{ij}}{\sum_{k\in S}
 C_{ik}},\quad C_{ij} := \#\{x_0\in B_i,\,
 x_T\in B_j\}. 
 \label{eq:P}
\end{equation}
Note that $\sum_{j\in S} P_{ij} = 1$ for all $i\in S$, so $P$ is a
row-stochastic matrix that defines a \emph{Markov chain} on boxes,
which represent the states of the chain.\cite{Bremaud-99, Norris-98}
The evolution of the discrete representation of $f(x)$, i.e., the
probability vector $\mathbf f = \smash{(f_i)_{i\in S}}$,
$\smash{\sum_{i\in S}} f_i = 1$, is calculated under \emph{left}
multiplication, i.e.,
\begin{equation}
 \mathbf f\mapsto \mathbf fP,
 \label{eq:pP}
\end{equation}
as it follows by noting that $\Pr(X_{(n+1)T}\in B_j) = \smash{\sum_{i\in
S}} \Pr(X_{(n+1)T}\in B_j, X_{nT}\in B_i) = \smash{\sum_{i\in S}}
\Pr(X_{nT}\in B_i) P_{ij}$. In this paper, whenever we multiply
vectors by matrices, we assume that the vector takes the appropriate
form of a row or column vector for the given operation.

Because $P$ is stochastic, $\mathbf 1 = (1, \dotsc, 1)$ is a
\emph{right} eigenvector with eigenvalue $\lambda = 1$, i.e.,
$P\mathbf 1 = \mathbf 1$. The eigenvalue $\lambda = 1$ is the largest
eigenvalue of~$P$. The associated potentially nonunique \emph{left}
eigenvector $\bpi = (p_i)_{i\in S}$ is invariant, because $\bpi P
= \bpi$ and can be chosen componentwise nonnegative (by the
Perron--Frobenius theorem).

We call $P$ \emph{irreducible} (or \emph{ergodic}) if for all $i,j\in
S$ there exists $n_{ij} \in\mathbb Z_0^+\setminus\{\infty\}$ such
that $\smash{(P^{n_{ij}})_{ij} > 0}$. To wit, all states of an
irreducible Markov chain communicate, the eigenvalue $\lambda
= 1$ is simple, and the corresponding left eigenvector $\bpi$ is
strictly positive.\cite{Norris-98} We call $P$ \emph{aperiodic} (or
\emph{mixing}) if there exists $i\in S$ such that $\gcd\{n\in\mathbb
Z_0^+ : \smash{(P^n)_{ii} > 0}\} = 1$. No state of an aperiodic
Markov chain is visited cyclically.

If $P$ is ergodic and mixing, then $\bpi$, normalized to a probability
vector ($\sum_{i\in S}p_i = 1$), satisfies $0 < \bpi = \bpi P =
\lim_{n\uparrow\infty}\mathbf f P^n$ for any probability vector
$\mathbf f$. We call $\bpi$ an invariant limiting probability vector
or \emph{stationary distribution}.

We adopt the traditional notation with $\smash{\{X_t\}_{t\in\mathbb
Z}}$ instead of $\{X_{nT}\}_{n\in\mathbb{Z}}$ and write, for instance, $P_{ij} =
\Pr(X_{t+1} = j\mid X_t = i)$, when this simplifies the notation.
In what follows we will assume that $P$ is both ergodic and mixing,
and the system is in stationarity, i.e., $\Pr(X_t\in B_i) = p_i$
for all $t\in \mathbb Z$. 

The Markov chain model we will deduce from data in Sec.\
\ref{sec:preparation_data} is, however, open, thus not ergodic.
For this reason, we shall next consider the closure of open dynamics.

\section{Closure of open dynamics}\label{sec:closure}

Let us assume that the flow domain is no longer closed, meaning
that trajectories can flow out of the domain and back into it. This
can happen for instance when the domain of interest is a subregion
of the closed world ocean domain $\mathcal X$ or when trajectory
data are only available in a subregion of $\mathcal X$. Other
possibilities include poor sampling of $\mathcal X$, weak communication
within, or the situation we describe in Sec.~\ref{sec:MCM}. In
every case the resulting dynamical system represents an \emph{open
dynamical system}.

The above is a slight variation of the setting in Sec.~\ref{sec:setup}.
We still assume that the motion is described
by a discrete-time-homogeneous Markov chain on a box covering
$\{B_i\}_{i\in O}$ of the ocean domain $\mathcal X$ but the probability
to transition from one box with index $i\in O$ to anywhere else in
the domain $O$ is no longer strictly $1$ since probability mass can
flow out of the domain. We denote the transition matrix on the open
domain by $P^O$ with entries given by $\smash{P^O_{ij}} := \Pr
(X_{t+1} = j\mid X_t = i)$ for $i,j\in O$. Since the rows of $P^O$
no longer have to add up to one, $\smash{P^O}$ represents a
\emph{substochastic} matrix.
  
We assume that a larger domain $S\supset O$ exists on which the
dynamics are closed, i.e., the transition matrix $P$ on box entries
$i,j \in S$ is stochastic. Furthermore, when we say that the dynamics
on the open domain is stationary, we actually mean that the dynamics
on the larger, closed domain is stationary with distribution $\bpi
= (p_i)_{i\in S}$, while we denote the restriction to the open
domain by~$\bpi \vert_O= (p_i)_{i\in O}$.

For further analysis it is often useful to artificially close the
open system. From the closure of $P^O$, we can, for instance, get
an estimate of $\bpi\vert_O$.  Closing $P^O$ can be done by appending
to $O$ a state $\omega$, which we will call \emph{two-way nirvana
state}, and letting all the outflow from $O$ flow into $\omega$,
while also redistributing the probability mass from $\omega$
\emph{back} into~$O$. Since thereby all boxes that are in $S$ but
not in $O$ are lumped together, this restricted dynamics should be
consistent with the original one under the assumption of well-mixedness
between exit from $O$ and reentry into it. For simplicity of notation,
we will denote the singleton $\{\omega\}$ also by $\omega$ and refer
to it too as the two-way nirvana state.

The resulting transition matrix on $O\cup \omega$ reads (possibly
overloading the notation by denoting it by $P$ again)
\begin{equation}
 P =
 \begin{pmatrix}
	 P^O & P^{O\to\omega}\\ 
	 P^{\omega\to O} & 0
 \end{pmatrix}
 \label{eq:closure}
\end{equation}
where $P^{O\to\omega} := \smash{\big(1 - \sum_{j\in O} P^O_{ij}\big)_{i\in
O}}$ (understood as a column vector) gives the outflow from $O$ to
$\omega$ and $\smash{P^{\omega\to O}}$ is a (row) vector that gives
the inflow and has to be a probability vector. Note that the matrix $P$ is
stochastic $\smash{\sum_{j\in O\cup \omega}} \smash{P_{ij}} = 1$
for all $i\in O\cup \omega$ and as such constitutes a \emph{closed
dynamical system}.

When no information about the reentry is available, e.g., because
data outside the open domain of interest are not available, a
possible choice \cite{Froyland-etal-14c} for $\smash{P^{\omega\to
O}}$ is to redistribute according to the quasistationary distribution
of $P^O$. \citet{Lunsmann-etal-18} alternatively use contour advection
for estimating the transition probabilities between boxes.  Without
adding a nirvana state, \citet{Froyland-etal-14c} immediately
redistribute the outflow back into the system. Here we redistribute
in such a way that accounts for ocean pollution, as we describe in
Sec.~\ref{sec:MCM}.

In the next section we will see how to study transitions between
$A$ and $B$ (subsets of $O$) in both the cases where i) the domain
is closed, i.e., $O=S$, and where ii) paths only traverse the open
domain $O \subsetneq S$. In the latter case, for the TPT computations
only knowledge of $P^O$ and the estimate of the stationary density
on the open computational domain $\bpi\vert_O$ is necessary.

\section{Transition Path Theory}\label{sec:tpt}

\subsection{TPT for closed systems}\label{ssec:tpt_closed}

\begin{figure}[tb]
 \centering%
 \includegraphics[width=.65\columnwidth]{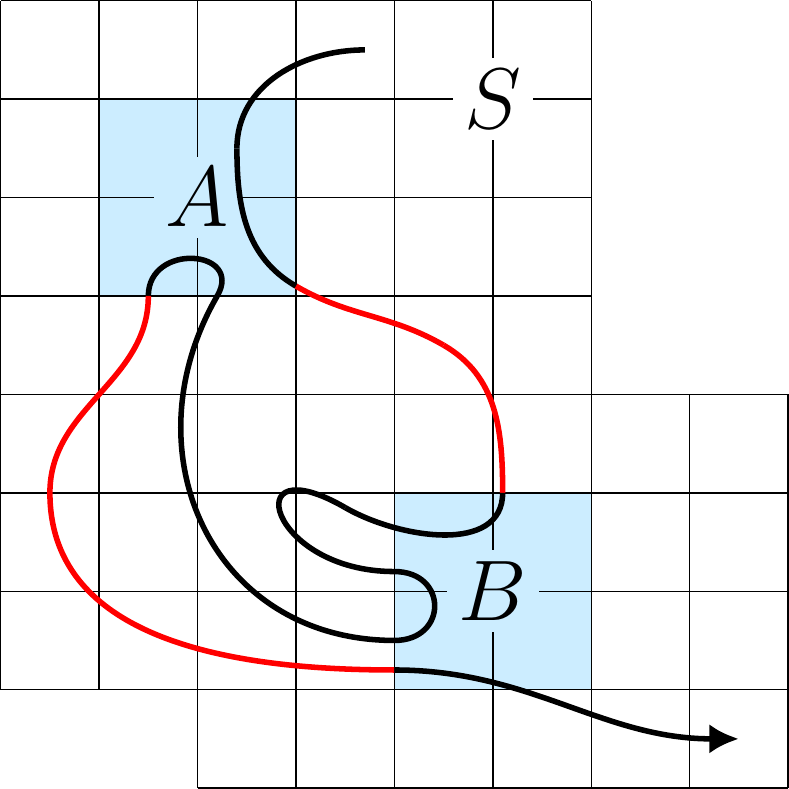}%
 \caption{Given a Markov chain taking values on $S$, the cartoon
 shows in red the reactive pieces of a trajectory connecting
 disjoint sets $A,B\subset S$.}
 \label{fig:tpt}%
\end{figure}

Motivated by a desire to understand rare events such as transformations
involved in chemical reactions, TPT provides a rigorous approach
to study transitions from a set $A\subset S$ to another, disjoint
set $B\subset S$ of a Markov chain. The results presented below
pertain to time-homogeneous (i.e., autonomous)
chains;\cite{VandenEijnden-06, Metzner-etal-06, Weinan-VandenEijnden-06,
Metzner-etal-09} extensions to the nonautonomous case have been
recently derived,\cite{Helfmann-etal-20} but they are beyond the
scope of this paper. Traditionally, source set $A$ is thought to
be formed by \emph{reactant} states, while target set $B$ of
\emph{product} states. Thus transitions from $A$ to $B$ are referred
to as \emph{reaction events}, while the pieces of trajectories
\emph{running from $A$ to $B$ without going back to $A$ or going
through $B$ in between} are known as \emph{reactive trajectories},
which are the focus of TPT (Fig.\ \ref{fig:tpt}).

The main tools of TPT are the forward and backward \emph{committor
probabilities} giving the probability of a random walker to hit $B$
before $A$, in either forward or backward time. The committor
probabilities are used to express various statistics of the ensemble
of reactive trajectories: i) the \emph{density of reactive
trajectories}, which provides information about the bottlenecks
during the transitions; ii) the \emph{current of reactive trajectories}
indicating the most likely transition channels; iii) the \emph{rate
of reactive trajectories} leaving $A$ or entering $B$; and iv) the
\emph{mean duration} of reactive trajectories. We will introduce
these in the following. Recall that we assume the chain to be
stationary with distribution $\bpi$.

The \emph{first entrance time} of a set $\mathcal S\subset S$ is
the stopping time random variable defined as 
\begin{equation}
  \tau^+_{\mathcal S} := \inf\{t \geq 0 : X_t\in \mathcal S\}   
\end{equation}
where $\inf\emptyset
:= \infty$. The \emph{forward committor} $\mathbf q^+ :=
\smash{(q_i^+)_{i\in S}}$ gives the probability that a trajectory
starting in $i\in \mathcal S$ \emph{first enters $B$, not $A$},
i.e., 
\begin{equation}
 q_i^+ : = \Pr(\tau^+_B < \tau^+_A \mid X_0 = i).
\end{equation}
Note that $\smash{q^+_{i\in A}} = 0$ while $\smash{q^+_{i\in B}} =
1$. For $i\in C:= S\setminus (A\cup B)$, one has that 
\begin{equation}
  q^+_i = \sum_{j\in S} P_{ij}q^+_j.
\end{equation}
The solution to this algebraic system is unique due to the
irreducibility of $P$, and in matrix notation expressed as
\begin{equation}
 \left\{
	 \begin{aligned}
	 \mathbf q^+\vert_C &= \big(\Id^{|C|\times |C|} -
	 P\vert_C\big)^{-1} P\vert_{C,B}\mathbf 1^{|B|\times
	 1},\\ 
	 \mathbf q^+\vert_A &= \mathbf 0^{|A|\times 1},\\ 
	 \mathbf q^+\vert_B &= \mathbf 1^{|B|\times 1},
  \end{aligned}
 \right.
 \label{eq:qp}
\end{equation}
where $\vert_{\mathcal S}$ denotes the restriction on indices in
$\mathcal S$, while $\vert_{\mathcal S,\mathcal S'}$ gives the
restriction to rows corresponding to $\mathcal S$ and columns of
$\mathcal S'$, if $\mathcal S = \mathcal S'$ we shorten this to
$\vert_{\mathcal S}$.

The \emph{last exit time}, in turn, is defined by 
\begin{equation}
 \tau^-_{\mathcal S} := \sup\{t\leq 0 : X_t\in \mathcal S\}
\end{equation} 
where $\sup\emptyset := -\infty$, which is a stopping time, but for
the \emph{time-reversed} chain $\{X^-_t\}_{t\in \mathbb Z}$ that
traverses the original Markov chain backwards in time, i.e. $X^-_t:=
X_{-t}$. The reversed chain's transition matrix, $P^- =
\smash{(P^-_{ij})_{i,j\in S}}$ is given by
\begin{equation}
 P^-_{ij} = \Pr(X_t = j \mid X_{t+1} = i) =
 \frac{p_j}{p_i}P_{ji},
\end{equation}
since the chain is assumed to be in stationarity. The time-reversed
transition matrix $P^-$ is ergodic and mixing, and has the same
stationary distribution $\mathbf p$ as $P$. The \emph{backward
committor} $\mathbf q^- : = \smash{(q_i^-)_{i\in S}}$ gives the
probability that a trajectory starting in $i\in \mathcal S$ \emph{last
exits $A$, not $B$}:
\begin{equation}
  q_i^- : = \Pr(\tau^-_A > \tau^-_B \mid X_0 = i).
 \end{equation}
In this case, 
\begin{equation}
  \smash{q^-_i} = \smash{\sum_{j\in S} P^-_{ij}q^-_j}
\end{equation}
for $i\in C$, subject to $\smash{q^-_{i\in B}} = 0$ and $\smash{q^-_{i\in
A}} = 1$. The (unique) solution in matrix notation,
\begin{equation}
 \left\{
	 \begin{aligned}
	 \mathbf q^-\vert_C &= \big(\Id^{|C|\times |C|} -
	 P^-\vert_C\big)^{-1} P^-\vert_{C,A}\mathbf 1^{|A|\times
	 1},\\ 
	 \mathbf q^-\vert_A &= \mathbf 1^{|A|\times 1},\\ 
	 \mathbf q^-\vert_B &= \mathbf 0^{|B|\times 1}.
  \end{aligned}
 \right.
 \label{eq:qm}
\end{equation} 
A particular situation arises in the special case when the chain
is reversible, namely, when $p_i P_{ij} =  p_j P_{ji}$ or, equivalently,
$P^- = P$. In such a case, $\mathbf q^- = \mathbf 1 - \mathbf q^+$.

The committors contain information that enable the computation of
various transition statistics. The \emph{distribution of reactive
trajectories} $\boldsymbol\mu^{AB} = \smash{(\mu^{AB}_i)_{i\in S}}$,
defined as the joint probability that the chain is in state $i$
while \emph{transitioning from $A$ to $B$}, viz.,
\begin{equation}
 \mu^{AB}_i := \Pr(X_0 = i, \tau^-_A > \tau^-_B,
 \tau^+_B<\tau^+_A),
\end{equation}
tells us where reactive trajectories spend most of their time.  Note
that $\smash{\mu^{AB}_{i\in A\cup B}} = 0$. The distribution of
reactive trajectories is computable from the committor probabilities
and the stationary distribution,
\begin{equation}
 \mu^{AB}_i = q^-_ip_iq^+_i.
 \label{eq:mu}
\end{equation}

A \emph{density of reactive trajectories}
$\smash{\boldsymbol{\hat\mu}^{AB}} = \smash{({\hat\mu}^{AB}_i)_{i\in
S}}$ is obtained by normalizing $\mu^{AB}_i$ by the probability to
be reactive
\begin{equation}
 Z^{AB} := \sum_{j\in C} \mu^{AB}_j = \Pr(\tau^-_A > \tau^-_B,
 \tau^+_B<\tau^+_A),
 \label{eq:Z}
\end{equation}
as it follows from the law of total probability. The result is
\begin{equation}
 \hat\mu^{AB}_i := \frac{\mu^{AB}_i}{Z^{AB}} = \Pr(X_0 = i\mid
 \tau^-_A > \tau^-_B, \tau^+_B<\tau^+_A),
\end{equation}
i.e., the probability of being in state $i$ conditioned on being
already on a reactive path from $A$ to $B$.

The \emph{current (or flux) of reactive trajectories} $\smash{f^{AB}} =
\smash{(f^{AB}_{ij})_{i,j\in S}}$ gives the average flux of
trajectories going through $i$ and $j$ at two consecutive times while
on their way from $A$ to $B$:
\begin{equation}
 f^{AB}_{ij} :=\Pr(X_0 = i,X_{1} =j, \tau^-_A > \tau^-_B,
 \tau^+_B<\tau^+_A),
\end{equation}
which is computable as
\begin{equation}
 f^{AB}_{ij} = q^-_ip_iP_{ij}q^+_j.
 \label{eq:f}
\end{equation}
Note that the reactive
current can include direct transitions from $i\in A$ to $j\in B$, which
are not accounted for in the corresponding reactive distribution
as it only considers transitions passing through $C$.

To eliminate
detours of reactive currents, one introduces the \emph{effective
current of reactive trajectories} $f^+ = (f^+_{ij})_{i,j\in S}$,
which gives the net amount of reactive current going through $i$
and $j$ consecutively, viz.,
\begin{equation}
 f^+_{ij} := \max\left\{f^{AB}_{ij} - f^{AB}_{ji},0\right\}.
 \label{eq:fp}
\end{equation}

To visualize $f^+$ on a flow domain covered by boxes $\{B_i\}_{i\in
S}$, one usually depicts the magnitude and the direction of the
effective current out of each $i$, i.e., to each $i$ one attaches
the vector $\smash{\sum_{j\neq i}f^+_{ij}e_{ij}}$, where $e_{ij}$
is the unit vector pointing from the center of box $B_i$ to the
center of $B_j$.  There also exists a flow decomposition algorithm
for extracting the dominant transition paths from
$f^+$.\cite{Metzner-etal-09}

The rate of transitions leaving $A$ or \emph{departure rate} is
defined as the probability per time step of a reactive trajectory
to leave $A$, i.e.,
\begin{equation}
 k^{A\to} := \Pr(X_0\in A, \tau^+_B<\tau^+_A) = \sum_{i\in A,
 j\in S} f^{AB}_{ij}
\end{equation}
and can be computed by summing up the reactive flux that exits $A$.
In turn, the rate of transitions entering $B$ or \emph{arrival rate}
is defined as the probability per time step of a reactive trajectory
to enter $B$:
\begin{equation}
 k^{B\leftarrow} := \Pr(X_0\in B, \tau^-_A>\tau^-_B) = \sum_{i\in
 S, j\in B} f^{AB}_{ij}.
\end{equation}
By a simple calculation, it can be shown that summing the reactive
current out of $A$, $\smash{\sum_{i\in A, j\in S}} \smash{f^{AB}_{ij}}$,
is equal to aggregating the reactive current into $B$, $\smash{\sum_{i\in
S, j\in B}} \smash{f^{AB}_{ij}}$, thus
\begin{equation}
 k^{A\to} = k^{B\leftarrow} =: \smash{k^{AB}}.
 \label{eq:kAB}
\end{equation}

To better interpret the transition rate $k^{AB}$, we give two
meanings. Consider an infinite $\bpi$-distributed ensemble of random
walkers in our domain, then at any time the proportion of random
walkers that are exiting $A$ while on their way to $B$ (or equivalently,
entering $B$ when coming last from $A$) is given by $k^{AB}$. Now,
on the other hand, consider only one random walker in the system,
then $k^{AB}$ can be interpreted as a frequency, i.e., the random
walker exits $A$ on average every $\smash{(k^{AB})^{-1}}$-th time
on the way to $B$ (and, equivalently, enters $B$ when coming from
$A$).

In some situations, e.g., when $B$ is given by a disconnected
set, it is insightful to further decompose the transition rate
\begin{equation}
  k^{B \leftarrow} = \sum_{B_n\subset B} k^{B_n\leftarrow} 
\end{equation}
into the individual arrival rates into  disjoint subsets $B_n$ that
together give $B = \cup_n B_n$:
\begin{equation}
k^{B_n\leftarrow} = \Pr(X_0\in B_n, \tau^-_A>\tau^-_B) = \sum_{i\in
 S, j\in B_n} f^{AB}_{ij}.
 \label{eq:decomp_rate}
\end{equation}
The same can also be done for decomposing $k^{A\to}$.

Finally, dividing the probability of being reactive by the discrete
transition rate,
\begin{equation}
 t^{AB} := \frac{Z^{AB}}{k^{AB}},
 \label{eq:tA}
\end{equation}
gives the \emph{expected duration} of a transition from $A$ to
$B$.\cite{VandenEijnden-06, Helfmann-etal-20}

We close this section with a remark on comparing probabilistic
computations with counting.  Ergodicity of the chain implies that
the objects in TPT can be approximated by ``counting'' transition
events of one sufficiently long trajectory, and this approximation
converges almost surely as the length of the trajectory tends to
infinity.\cite{VandenEijnden-06, Helfmann-etal-20} For instance,
the forward committor $q_i^+$ of any state $i$ is approximated by
the fraction of all visits of the chain to state $i$ after which
the chain directly transitioned to $B$ without hitting $A$ first.
All other quantities considered here can be similarly approximated.
As we intend to apply TPT to a chain extracted from drifter trajectory
data, one might wonder whether this level of sophistication is
necessary to our ends or whether one could simply do an  approximation
by counting. The answer lies in the features of the data.  One would
need sufficiently many drifter trajectories that are sufficiently
long to resolve the transition statistics, and that are also spread
according to the right distribution. None of these requirements are
met, and the best one can do is to ``concatenate'' the drifter
information into a Markov chain, as it will be done in Sec.~\ref{sec:MCM}
below.

\subsection{TPT for open domains}\label{ssec:tpt_open}

To apply TPT to open dynamical systems on $O$, a modification from
the standard setting as reviewed in Sec.~\ref{ssec:tpt_closed} is
needed.  Adding the state $\omega$ to $O$ closes the system
artificially (as in Sec.~\ref{sec:closure}) but we are still only
interested in the transitions from $A\subset O$ to $B\subset O$
that \emph{stay} in $O$ during the transition. Thus the reactive
trajectories we consider go from $A$ to $B$ without passing $A$,
$B$ or $\omega$ during the transition. If we were to apply the usual
TPT on the artificially closed system we would also observe artificial
transitions via the added state~$\omega$.

In order to compute the statistics of the reactive trajectories
from $A$ to $B$ only through $O$ we look at slightly different
committors. Namely, the forward committor now gives the probability
to next transition to $B$ rather than to $A$ or outside of $O$ when
starting in state $i$, i.e.,
\begin{equation}\label{eq:q+_open}
 q_i^+ : = \Pr(\smash{\tau^+_B} < \tau^+_{A\cup\omega} \mid X_0
 = i),
\end{equation}
while the backward committor gives the probability to have last
come from $A$, not $B\cup\omega$
\begin{equation}\label{eq:q-_open}
 q_i^- : = \Pr(\smash{\tau^-_A} > \tau^-_{B\cup\omega} \mid X_0
 = i).
\end{equation}
In that way the product of forward and backward committors becomes
the probability when initially in $i$ to have last come from $A$
and next go to $B$ while not passing through $A$, $B$ or $\omega$
in between.

By definition, the forward committor is $q^+_i = 0$ for $i\in
A\cup\omega$ and $1$ for $i\in B$, while in the transition region
$C := O\setminus (A\cup B)$ it satisfies
\begin{equation}
 q^+_i = \sum_{j\in O\cup\omega}P_{ij}q^+_j = \sum_{j\in
 O}P_{ij}q^+_j + P_{i\omega}q^+_{\omega} = \sum_{j\in O}P^O_{ij}q^+_j
\end{equation}
since $q_{\omega}^+ = 0$ and $P$ on entries of $O$ reduces to $P^O$.

The backward committor $q^-_i = 0$ for $i\in B$ and 1 for $i\in
A\cup\omega$, while, by a similar reasoning as above, it satisfies
\begin{equation}
 q^-_i = \sum_{j\in O}P^{O,-}_{ij}q^-_j
\end{equation}
for $i\in C$, where $P^{O,-}$ is the restriction of the backward-in-time
transition matrix $P^-$ to $O$ and has entries $P^{O,-}_{ij} =
\frac{p_j}{p_i}P^O_{ji}$ for $i,j\in O$.

Therefore, system \eqref{eq:qp} remains the same with the replacement
of $P$ with $P^O$ and $A$ with $A\cup\omega$. In turn, system
\eqref{eq:qm} remains the same with the replacement of $P^-$ with
$P^{O,-}$ and $B$ with $B\cup\omega$.

The rest of the formulae in Sec.~\ref{ssec:tpt_closed} are not
changed except that the committors are now given as above. An
important observation, however, is that $\mu^{AB}_i = 0$ for $i=
\omega$ and $\smash{f^{AB}_{ij}} = 0$ for $i,j=\omega$. Thus only
their values on $O$ are of interest, where $P$ can be replaced by
$P^O$ and $\mathbf p$ can be substituted by its restriction to $O$,
$\mathbf p\vert_O$. Also, as the rate and mean transition time of
reactive trajectories are derived from the density and current,
they are computable solely from $P^O$ and $\mathbf p\vert_O$.

This version of TPT for open dynamics, can also be useful in other
settings, e.g., when one wants to study transitions between $A$ and
$B$ that avoid a third subset $D$ of the state space $S$.

\section{Markov-chain model for ocean pollution}\label{sec:MCM}

In the following we describe our stochastic model for the dynamics
of a single plastic debris piece that enters the ocean at the coast
with a probability reflecting observed levels of mismanaged plastic
waste in near coastal communities.  From the coast, the debris piece
traverses the ocean, possibly passing and staying for long times
near garbage patches.  Its motion is fitted using satellite-tracked
drifter trajectories; cf.\ Sec.~\ref{sec:preparation_data}.  Whenever
a debris piece beaches somewhere, we reinject it again next to the
coast. The coastal injection and beaching is described in
Sec.~\ref{sec:pollution-aware} below.

In that way we will model the behavior of a generic plastic debris
piece in the ocean by a stationary ergodic Markov chain.  Of course,
there is a huge amount of plastic debris in the ocean, each day
growing in number.  But we are not interested in modeling the change
in plastic concentration in the ocean.  Rather, our interest lies
in understanding the routes of plastic waste from the coasts to the
garbage patches by means of a TPT analysis. This distinction is
elaborated on in Sec.~\ref{ssec:physical_interpret}.

\subsection{Preparation of $P$ from drifter trajectory data}\label{sec:preparation_data}

As anticipated, to formulate the Markov chain for marine debris
motion we use drifter trajectory data, taken from the NOAA Global
Drifter Program.\cite{Lumpkin-Pazos-07} Satellite-tracked by the
\emph{Argos} system or GPS (Global Positioning System), the drifters
from this database have a spherical surface float with a 15-m-long
holey-sock drogue attached.\cite{Sybrandy-Niiler-91} They are
engineered to resist wind slippage and wave-induced drift, and hence
to follow water motion as close as possible.\cite{Niiler-Paduan-95}
We therefore only consider trajectory portions during which the
drifter's drogue has been lost,\cite{Lumpkin-etal-12} which can be
expected to provide a more fair representation of floating marine
debris motion.\cite{Beron-etal-16, Beron-etal-19-PoF, Olascoaga-etal-20,
Miron-etal-20-PoF, Beron-20}

The basic procedure to construct the transition matrix $P$, defined
in \eqref{eq:P}, is as follows. We first interpolate the available
undrogued drifter trajectories daily and form two arrays, one
representing positions at any instant of time over 1992--2019 ($x_0$)
and another one representing their images ($x_T$) after $T = 5$ d.
Here we are assuming that the ocean motion did not change considerably
over the last 30 yr such that the transition matrix $P$ from this
data set is still a good representation of the ``average ocean
motion.''

We then define the box covering $\smash{\{B_i\}_{i\in S}}$ by lying
down on the world ocean domain a grid of roughly 3$^{\circ}$ width
(due the planet's curvature the area of the boxes is not fixed,
varying from 100--10000 km$^2$, but this is inconsequential in the
definition of the vector space $V_N$, normalized by box area). The
entries of $P$ are finally estimated via counting according to
\eqref{eq:P}. As in previous work \cite{Miron-etal-17, Olascoaga-etal-18,
Miron-etal-19-JPO, Miron-etal-19-Chaos, Beron-etal-20-Chaos} the
transition time $T$ is chosen long enough to guarantee negligible
memory into the past and sufficient communication among boxes, made
large enough to maximize sampling. The simple Markovianity test
$\lambda(P(nT)) = \lambda(P(T))^n$ is passed well up to $n = 4$.

There are additional aspects, not encountered, partially addressed
or overlooked earlier, which must be coped with to make $P$ meaningful.
\begin{enumerate}
 \item \emph{Zonal connectivity}. This is addressed by identifying
  and continuating trajectories crossing the antimeridian connecting
  the eastern and western hemispheres.
\item \emph{Spurious communication between ocean basins}. This
  situation occurs where ocean basins are separated by narrow land
  masses.  The situations that concern us are the Panama Isthmus
  separating the Pacific and Atlantic Ocean basins, and also the
  Maritime Continent separating the Pacific and Indian Oceans.
  Neither the undrogued drifters considered nor drogued drifters
  analyzed earlier \cite{McAdam-vanSebille-18} reveal connectivity
  between the Pacific and Indian Oceans through the various straits
  and passages in that region, which might seem at odds with the
  presence of the Indonesian Throughflow,\cite{Gordon-Fine-96}
  particularly for the drifters drogued at 15 m.  However, this
  takes place mainly within the thermocline layer (50--200 m),
  \cite{Tillinger-Gordon-09} which is less correlated with local
  wind flow that quite strongly affects the undrogued drifters and
  also the drogued drifters, albeit to a lesser extent.  To avoid
  spurious communication between the basins we proceed as follows.
  Let $B_k$ be a box spanning portions of for instance the Pacific
  Ocean and Atlantic Ocean (Caribbean Sea).  Denote $B_k^\mathrm{PO}$
  and $B_k^\mathrm{AO}$ the portions of $B_k$ lying on the Pacific
  Ocean and Atlantic Ocean sides, respectively.  In computing
  transitions between $B_k$ and other boxes we only consider those
  from or into $B_k^\mathrm{PO}$ or $B_k^\mathrm{AO}$ depending on
  which one makes the largest number of transitions.  This guarantees
  that $P_{kj} > 0$ and $P_{jk} > 0$ exclusively for $j\in S$ such
  that $B_j$ is either in the Pacific Ocean or the Atlantic Ocean.
\item \emph{Nonobserved communication}. A prominent example
  of this is the communication between the Atlantic Ocean and
  Mediterranean Sea. Depending on the size of the boxes $B_i$, a
  connection might exists through the Gibraltar Strait, even thought
  in reality no drifter is seen to traverse it (in any direction).
  We resolve this situation by excluding the Mediterranean Sea
  domain from consideration.
 \item \emph{Weak communication}. We enable as much communication
  as possible along the chain by restricting the chain to the largest
  strongly communicating class of states. This is done by applying
  the Tarjan algorithm \cite{Tarjan-72} on the directed graph
  equivalent to the Markov chain. This procedure excludes boxes
  from the partition. Among those boxes are 22 poorly sampled coastal
  boxes, mainly in the Kara Sea of the Arctic Ocean and the Seas
  of Indonesia, with trajectories flowing in, but not flowing out
  in the next step.  Let $O$ be the ordered set of box indices in
  the largest class of strongly connected boxes.  Using the notation
  in Sec.~\ref{sec:closure}, we call $P^O$ the substochastic
  transition matrix characterizing this open system.  The Markov
  chain is now substochastic, since by the exclusion of boxes it
  is no longer ensured that probability mass is conserved.
\end{enumerate}

\subsection{Pollution-aware model derivation}\label{sec:pollution-aware}

To formulate our Markov-chain model for ocean pollution, we leverage
the possibility that marine debris get stuck at shorelines.  This
creates additional outflow of the system that must be compensated
for, which we choose to do in such a way as to model ocean pollution
at the coasts.

\begin{figure}[tb]
 \centering%
 \includegraphics[width=\columnwidth]{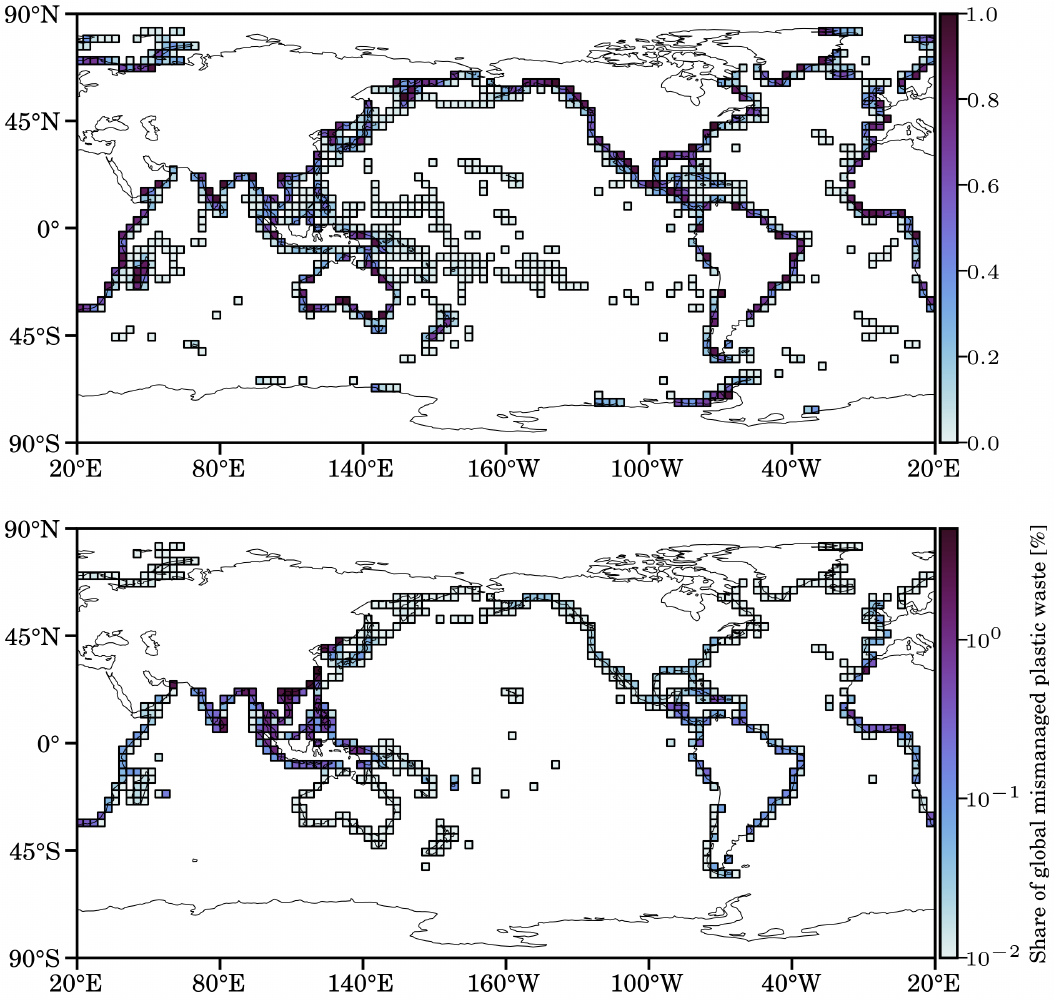}%
 \caption{(top panel) Fraction of land (or ice) filling coastal
 boxes of the surface world ocean partition (black). (bottom panel)
 Percentage of share of global mismanaged littered or inadequately
 disposed plastic waste estimated in 2010 for populations living
 within 50 km of the coastline.}
 \label{fig:ell}%
\end{figure}

Specifically, let $\ell : O \to [0,1)$ be a \emph{land fraction}
function giving the ratio between land area and total box area.
Namely, $0< \ell(i) <1$ for $i\in L\subset O$ corresponding to boxes
filled with some portion of land (or ice) (Fig.\ \ref{fig:ell}, top
panel) and $\ell(i) = 0$ otherwise. We then follow
\citet{Miron-etal-19-Chaos} and replace
\begin{equation}
  P^O_{ij} \gets \big(1-\alpha\ell(i)\big) P^O_{ij},\quad
  0\le \alpha \le 1
  \label{eq:PO}
\end{equation}
for all $i,j\in O$. To wit, only a fraction of the probability mass,
proportional to the amount of land covering box $B_i$, is allowed
to flow from $i$ to $j$, the remaining probability mass is assumed
to beach and flows out of the system.  The factor $\alpha$, not
considered in \citet{Miron-etal-19-Chaos}, was included to enable
consistency with observations.  While we have performed optimizations
of no kind, we have found that $\alpha = \smash{\frac{1}{4}}$
produces results most consistent with them.  If $\alpha = 1$ (as
in \citet{Miron-etal-19-Chaos}) then the so-called Great Pacific
Garbage Patch \cite{Kubota-94} in the North Pacific subtropical
gyre is not revealed as intense as observations
indicate.\cite{Cozar-etal-14, Lebreton-etal-18} However, transition
channels into this patch and patches in the other subtropical gyres
are not sensitive to the specific $\alpha$-value assumed, as we
show in the supplementary material.

To deal with the created substochasticity by a closure of the system,
we augment the chain by one artificial state $\omega$ as in
\eqref{eq:closure}.  All the outflow of the open system goes into
$\omega$ and we reinject the probability mass from $\omega$ to $O$
through coastal boxes according to plastic waste input from land
into the ocean, viz.,
\begin{equation}
  P^{\omega\to O}_i =
  \begin{cases}
	 \dfrac{W_i}{\sum_{i\in L} W_i}  & \quad \text{if } i\in L,\\
	 0  & \quad \text{otherwise}.
  \end{cases}
  \label{eq:PomegaO}
\end{equation}
Here $W_i$ is the mass of mismanaged plastic waste in $B_i$, $i\in
L$, as inferred from estimates \cite{Jambeck-etal-15} made in 2010
for populations living within 50 km of the coastline.  This is shown
in percentage of the total mass in the bottom panel of Fig.\
\ref{fig:ell}; note that only inhabited coastal boxes for which
estimates are available are shown.   We denote the transition matrix
on the closed domain by $P$, but it should not be confused with the
transition matrix $P$ from the above Sec.\ \ref{sec:preparation_data}
which has a different domain and entries.  The two-way nirvana state
$\omega$ compensates for the substochasticity of $P^O = P\vert_O$
by sending back into the chain any imbalances through the land
states distributed according to the ability of such states to
``pollute the oceans'' as inferred by their share of the global
land-based plastic waste entering the ocean through them. It must
be realized that in this \emph{statistical} model debris mass is
neither created nor destroyed.  In other words, the model assumes
that the world ocean is polluted by plastic at a certain level, and
that the ocean currents and winds redistribute the existing pollutants
within ocean basins.  If beaching occurs, then the pollutants are
returned back to the ocean in an equal quantity simulating mismanaged
plastic waste loading from land runoff.

\subsection{Physical interpretation of the model}
\label{ssec:physical_interpret}

We model the distribution of garbage input per time unit to the
oceans by a time-independent vector $P^{\omega\to O} =: \mathbf{r} \in
\mathbb{R}^{|O|}$. Each entry of $\mathbf{r}$   accounts for the
probability per time unit of injecting a garbage particle into the
corresponding box. Thus $\mathbf{r}$ is supported on the coastal
(land) boxes, i.e., $r_i =0$ for $i\notin L$, and $\mathbf{r}$ is
a probability vector, i.e., $\sum_i r_i=1$.

Then the total accumulated garbage \emph{mass distribution} in the
oceans is time-asymptotically going to be
\begin{equation}\label{eq:asymp_mass}
\sum_{k=0}^{\infty} \mathbf{r}  (P^O)^k = \mathbf{r} \, (\Id-P^O)^{-1}.
\end{equation}
Recall that $P^O$ is assumed to be irreducible, thus $\Id-P^O$ is
invertible.  Equation~\eqref{eq:asymp_mass} gives the  mass
distribution of debris particles entered over an infinite time
frame, thus it does not need to be a probability vector. It is the
limiting (saturated) mass distribution of pollution measured in the
units dictated by~$\mathbf{r}$.  If we would like to know the
relative distribution of garbage that has accumulated over time,
we would norm this vector to a probability vector.

Now, it turns out that the very same long-term distribution is
modeled by our ``recirculating'' Markov chain. With $\mathbf{a} :=
P^{O\to\omega}$ being the vector of absorption probabilities from
the boxes into nirvana (the outflow), the stationary distribution
of our chain satisfies
\begin{equation}
\begin{pmatrix}
\bpi \vert_O & \rho
\end{pmatrix}
\begin{pmatrix}
P^O& \mathbf{a}\\
\mathbf{r}  & 0
\end{pmatrix} = 
\begin{pmatrix}
\bpi \vert_O & \rho
\end{pmatrix},
\end{equation}
with stationary vector $\bpi \vert_O$ on indices of $O$ and scalar
$\rho$ giving the stationary weight of~$\omega$. The set of
equations corresponding to the boxes in $O$ read as $\bpi \vert_O
\, P^O + \rho \, \mathbf{r}  = \bpi \vert_O$, or, after rearrangement,
\begin{equation}
  \bpi \vert_O \, (\Id-P^O) = \rho \, \mathbf{r}.  
\end{equation}
Since $\rho$ is scalar, this readily means that $\bpi \vert_O \propto
\mathbf{r} \, (\Id-P^O)^{-1}$, which equals the asymptotic mass
distribution \eqref{eq:asymp_mass} from above. In summary, our
stationary Markov chain constructed with reinjection is the statistical
equivalent of garbage motion in the ocean, based on the limiting
garbage distribution.

\section{Long-time asymptotics}\label{sec:asym}

\begin{figure}[tbh]
 \centering%
 \includegraphics[width=\columnwidth]{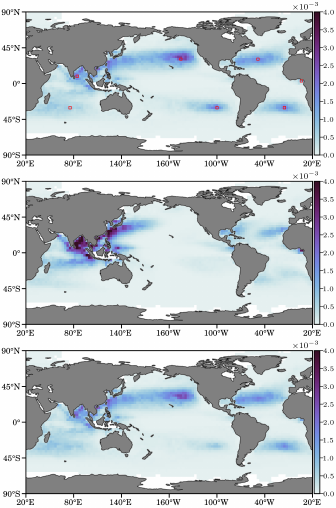}
 \caption{(top panel) Restricted to the set $O$ of boxes covering
 the physical ocean domain where the dynamics are open, the stationary
 distribution $\mathbf p$ of the closed dynamics represented by the
 transition matrix $P$ for marine debris pollution. Note that
 $\mathbf p$ locally maximizes inside the subtropical gyres, which,
 at the same time happen to develop great patches. Indicated by the
 red boxes are these (and additional; cf.\ text for details) local
 maxima of $\mathbf p\vert_O$. (middle panel) Restricted to $O$,
 distribution after 1-yr evolution under left multiplication by $P$
 of a probability density (vector) with support on the virtual
 nirvana state included to close the system. (bottom panel) As in
 the middle panel, but after 10 yr.}
 \label{fig:p}%
\end{figure}

By design, the proposed transition matrix $P$ for marine debris
pollution has a single maximal communicating class of the states,
implying irreducibility for $P$ and ergodicity for the dynamics.
Furthermore, direct pushforward (i.e., evolution under left
multiplication by $P$) of an arbitrary probability vector reveals
convergence to the dominant left eigenvector $\mathbf p$ (the chain
is also aperiodic), which is invariant and also limiting, and hence
represents a stationary distribution.  The top panel of Fig.\
\ref{fig:p} shows $\mathbf p > 0$ restricted to $O$, viz., the set
of boxes of the world ocean partition where the dynamics are open.
The middle and bottom panels show, restricted to $O$, the distribution
after 1 and 10 yr of evolution under left multiplication by $P$ of
$\mathbf 1_\omega$, respectively.  Note that $\mathbf p\vert_O$
locally maximizes in the subtropical gyres, quite evidently in the
eastern side of the North Pacific gyre. In most of the Indian Ocean
$\mathbf p\vert_O$ reveals several well-spread local maxima consistent
with a predominantly uniform distribution.  The exception is the
Bay of Bengal, where $\mathbf p\vert_O$ shows more clear sings of
local maximization. An additional local maximum of $\mathbf p\vert_O$
is seen in the Gulf of Guinea south of West Africa.  The several
local maxima of $\mathbf p\vert_O$ identified are indicated by the
red boxes in the top panel of Fig.\ \ref{fig:p}.  The Indian Ocean
location corresponds to its local maximum inside the subtropical
gyre.

\begin{figure}[tb]
 \centering%
 \includegraphics[width=\columnwidth]{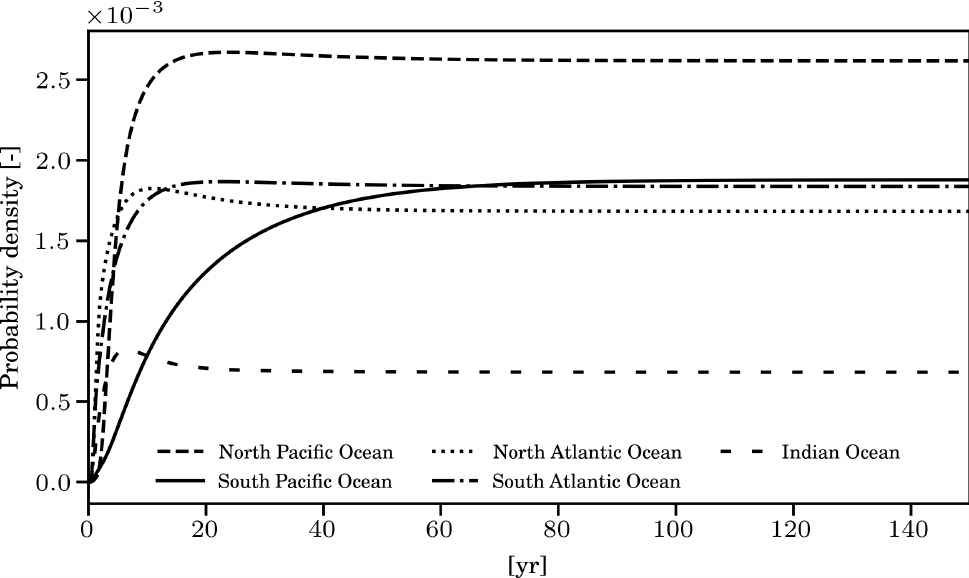}%
 \caption{Evolution of $\mathbf 1_\omega$
 under left multiplication by $P$ restricted to the boxes where
 $\mathbf p\vert_O$ locally maximizes in the subtropical gyres.}
 \label{fig:1APk}%
\end{figure}

The structure of $\mathbf p\vert_O$ suggests garbage patches in the
subtropical gyres of the Atlantic and Pacific Oceans consistent
with in-situ microplastic concentration observations.\cite{Cozar-etal-14}
Previous analyses \cite{Maximenko-etal-12, vanSebille-etal-12} of
drifter data revealed these patches too, albeit from direct evolution
of probability densities. In particular, \citet{vanSebille-etal-12}
argued that the North Pacific patch should be the main attractor
of global marine debris, in agreement with direct observational
evidence \cite{Lebreton-etal-18, Cozar-etal-14} of the Great Pacific
Garbage Patch.\cite{Kubota-94} Our pollution-aware model produces
consistent results. This can be anticipated from $\mathbf p\vert_O$
acquiring larger values in the North Pacific gyre than in the other
subtropical gyres, and also from direct pushforward of $\mathbf
1_\omega$ and subsequent restriction of the evolved density to the
boxes where $\mathbf p\vert_O$ locally maximizes in the subtropical
gyres (Fig.\ \ref{fig:1APk}).  (It should be noted too that the
structure of $\mathbf p\vert_O$ in the North Pacific suggests a
garbage patch, albeit weaker, in the western side of the basin in
agreement with field sampling.\cite{Yamashita-Tanimura-07}) The
relative weakness of the Indian Ocean garbage patch \cite{Cozar-etal-14}
attributed to unique oceanic and atmospheric dynamics in the region
\cite{vanderMheen-etal-19} is consistent with the results from our
Markov-chain model for ocean pollution too.  There the stationary
distribution $\mathbf p\vert_O$ does not reveal a clear local maximum
(Fig.\ \ref{fig:p}), and the direct pushforward of $\mathbf 1_\omega$
identifies the Indian Ocean gyre as the less attracting of all the
subtropical gyres (Fig.\ \ref{fig:1APk}).  Exactly where the garbage
patches are located is determined by wind-induced Ekman and
wave-induced Stokes drift effects \cite{Maximenko-etal-12} mediated
by the inertia (i.e., buoyancy and size) of the floating debris
pieces.\cite{Beron-etal-16, Beron-etal-19-PoF, Olascoaga-etal-20,
Miron-etal-20-PoF, Miron-etal-20-GRL, Beron-20}  Indeed, the numerical
simulations of inertial particles by \citet{Beron-etal-16} do not
reveal signs of accumulation in the Indian Ocean gyre as clear as
in the other gyres.

In \citet{vanSebille-etal-12} the authors suggest the possibility
of a rather strong garbage patch in the Barents Sea in the Arctic
Ocean, possibly constrained by slow surface convergence due to
deep-water formation.  While the authors noted that this patch might
be an artifact of drifters becoming grounded in the (seasonal)
sea-ice, observational support of plastic accumulating in the region
is emerging.\cite{Cozar-etal-17}   However, the observed accumulation
represents a very small fraction (3\pct) of the global standing
stock.  Our pollution-aware model does not reveal a patch there,
more consistent with this observation.

Our model suggests the occurrence of a patch in the Gulf of Guinea,
which seems to be supported only on numerical
simulations.\cite{Mountford-Morales-19} However, the the Gulf of
Guinea is identified as a mesopelagic niche with genomic characteristics
than different than its surroundings.\cite{Sutton-etal-17} This
patch remained elusive to earlier studies.\cite{Maximenko-etal-12,
vanSebille-etal-12} A likely explanation is the involvement in those
earlier studies of both undrogued \emph{and} drogued drifters, which
unlike floating debris, are much less affected by inertial
effects.\cite{Miron-etal-20-GRL}  However, a more recent study
\cite{vanderMheen-etal-19} involving exclusively undrogued drifters
did not reveal accumulation in the Gulf of Guinea time asymptotically.

The structure of $\mathbf p\vert_O$ also reveals that the Bay of
Bengal has potential for holding a garbage patch. High plastic
concentration in the Bay of Bengal has been reported and attributed
to loading from nearby land-based sources.\cite{Ryan-13} The
occurrence of a garbage patch in the Bay of Bengal was also suggested
recently from the analysis of undrogued drifter trajectory
data.\cite{vanderMheen-etal-19}

The pertinent question is how the garbage patches are filled. We
address this using TPT.

\section{Reactive debris paths}\label{sec:debris_paths}

With the above in mind, we proceed to apply TPT to the dynamics on
the physical world ocean domain, where reactive debris currents are
sought to be unveiled. The usual TPT (Sec.~\ref{ssec:tpt_closed})
allows us to compute statistics of the ensemble of reactive paths
of marine debris into garbage patches, with the help of TPT for
open domains (Sec.~\ref{ssec:tpt_open}) we can study reactive paths
between garbage patches.

\subsection{Pollution paths into garbage patches}

\begin{figure*}[tb]
 \centering%
 \includegraphics[width=\textwidth]{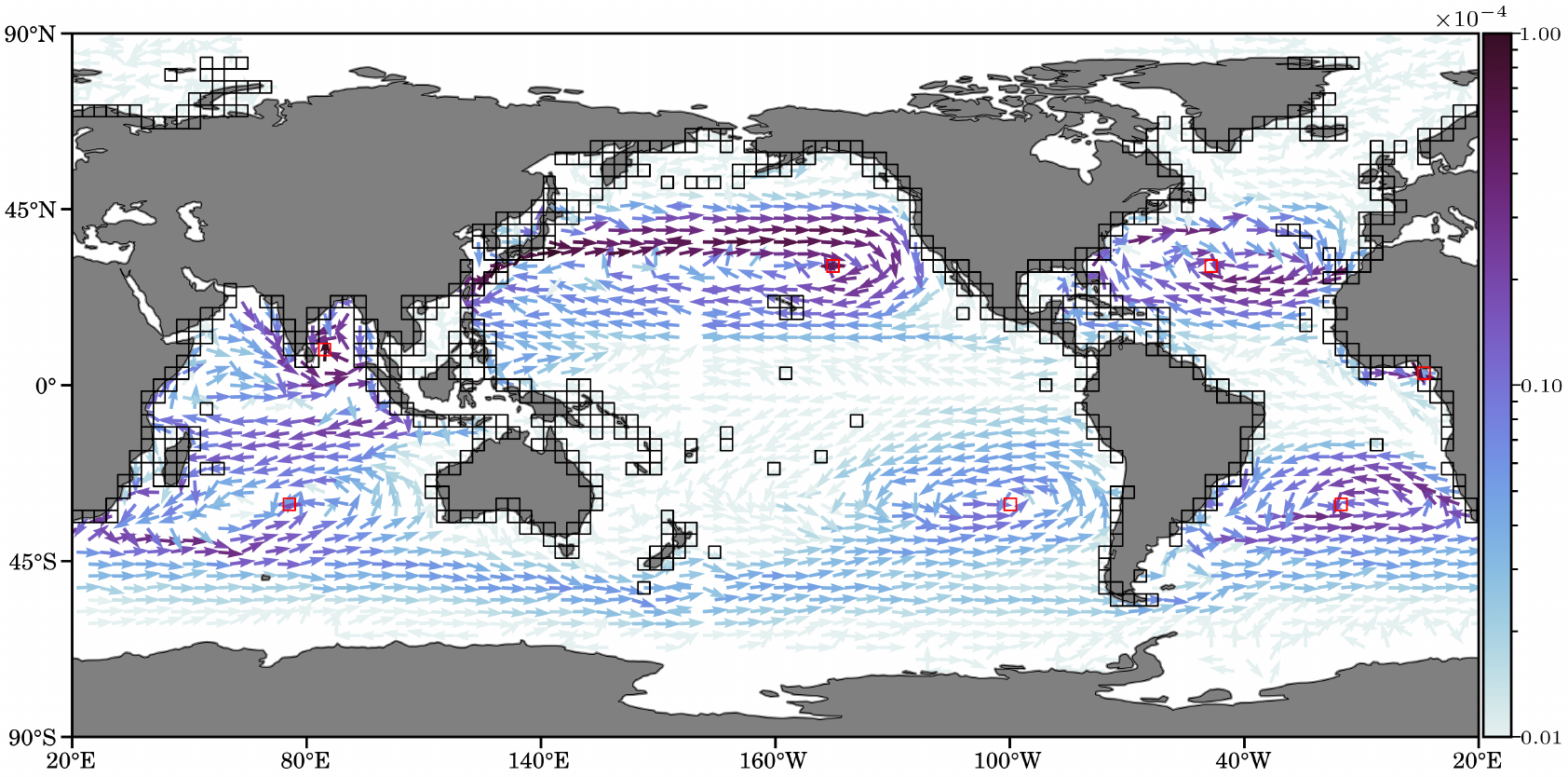}%
 \caption{Inferred reactive probability currents of marine debris
 into garbage patches (red boxes).  Black boxes indicate coastal
 boxes from which those currents emerge.}
 \label{fig:f}%
\end{figure*}

To infer the pollution paths into garbage patches, we choose the
nirvana state $\omega$ as the  source state $A$ of garbage, and we
identify the set of target states $B$ with the union of indices of
boxes covering garbage patches as inferred by the regions where
$\mathbf p\vert_O$ tends to locally maximize, which we have isolated
above (cf.\ Fig.\ \ref{fig:p}, top panel). We will denote $G$ the
garbage patch set. Although the debris is reentering the ocean
through the land boxes $L$, choosing the source as $A = \omega$ is
more reasonable, as it allows reactive debris trajectories to enter
boxes in $L$ and to flow on towards $B$. With the choice $A=L$ we
would have excluded this possibility, which would have caused a
notable impact on TPT computations, given the size of our boxes.
The effective currents of reactive trajectories resulting from the
TPT analysis are depicted in Fig.\ \ref{fig:f}, with the target set
$B = G$ indicated by the red boxes.  In black we depict the subset
of pollution-capable coastal boxes $L$.

We first note that the extent of the reactive currents running into
the subtropical gyre patches is in general larger than those running
into the near coastal patches.  These indicates that the near coastal
patches are mainly fed from nearby land-based plastic waste sources.
An exception is the Bay of Bengal patch, which appears to accumulate
garbage from remote sources in the coasts of the Arabian Sea and
even more remote ones in the coasts of Indonesia.  Particularly
constrained seems to be patch in the Gulf of Guinea, which is
inferred to be filled with plastic debris releases at the southern
coasts of West Africa. A refined assessment of these mostly qualitative
conclusions is presented below.

Continuing with the visual inspection of Fig.\ \ref{fig:f}, for the
North Pacific patch TPT analysis infers a robust zonal eastward
reactive channel into it straight out from the coasts of China.  A
good deal of the transported debris is inferred to travel back to
the western side of the North Pacific basin, where the stationary
distribution $\mathbf p\vert_O$ also tends to maximize. A pollution
source for the South Pacific patch is not restricted to the east
coast of South America. Indeed, a westerly transition channel
originating in the Indian Ocean and the coasts of New Zealand is
also identified. Two clear reactive paths into the North Atlantic
are identified, one mainly coming from the southeastern coast of
the United States and another one coming from the northern coasts
of West Africa. In turn, the South Atlantic patch is fed from debris
transport from the Brazil--Malvinas Confluence and the southern tip
of Africa. A main carrier of pollution for the Indian Ocean patch
is the Agulhas Return Current.  However, this pollution channel
bifurcates a bit east of the patch's longitude, where a branch
originates to ultimately feed the South Pacific patch.  Indeed, the
pattern of the currents near the Indian Ocean patch does not suggest
as clear channels into it as into the patches in the other subtropical
gyres.  This seems consistent with the reported \cite{vanderMheen-etal-19}
weakness of the Indian Ocean patch.

It is important to note that while ocean currents play a dominant
role in transporting debris, the reactive paths inferred by TPT do
not resemble entirely the mean surface-ocean currents. However,
this is not unexpected given the various mechanisms, noted above,
controlling the motion of floating material beyond advection by
ocean currents. We stress again that TPT, by construction, highlights
currents composed of only trajectories that go from $A$ (source)
to $B$ (target), thereby excluding information about currents that
go from $B$ to $A$, $A$ to $A$, and $B$ to $B$.

The expected transition duration \eqref{eq:tA} is estimated to be
2.6 yr from the coasts into the subtropical gyre patches and the
gulf and coastal sea patches. Note that this is the mean time a
reactive trajectory takes from being injected into the oceans to
hit \emph{any} of these patches. If we set $B = Y$, where $Y\subset
G$ is the set of indices corresponding to the subtropical gyre
patches, then the mean duration is 5.6 yr, cf.\ \eqref{eq:decomp_rate}
and~\eqref{eq:tA}. The expected durations of individual transition
paths into the North Pacific, South Pacific, North Atlantic, South
Atlantic, and Indian Ocean patches are 7.3, 8.6, 4.3, 4.0, and 4.2
yr, respectively. The mean durations of those into the patches in
the Bay of Bengal and the Gulf of Guinea are 0.6 and 0.2 yr,
respectively. The proximity to the coasts explain the short mean
durations of the latter transition channels. As for the transition
channels into the subtropical gyre patches, those into the South
Pacific and North Atlantic patches stand out as the overall slowest
and fastest in the class, respectively. These times to individual
patches represent the mean duration of those reactive trajectories
that first hit the set $B = g\in G$ through the respective patch,
i.e., transitions are direct and not through other patches, $G\setminus
g$ (which can be avoided by using TPT for open dynamics).

\begin{figure}[t]
 \centering%
 \includegraphics[width=\columnwidth]{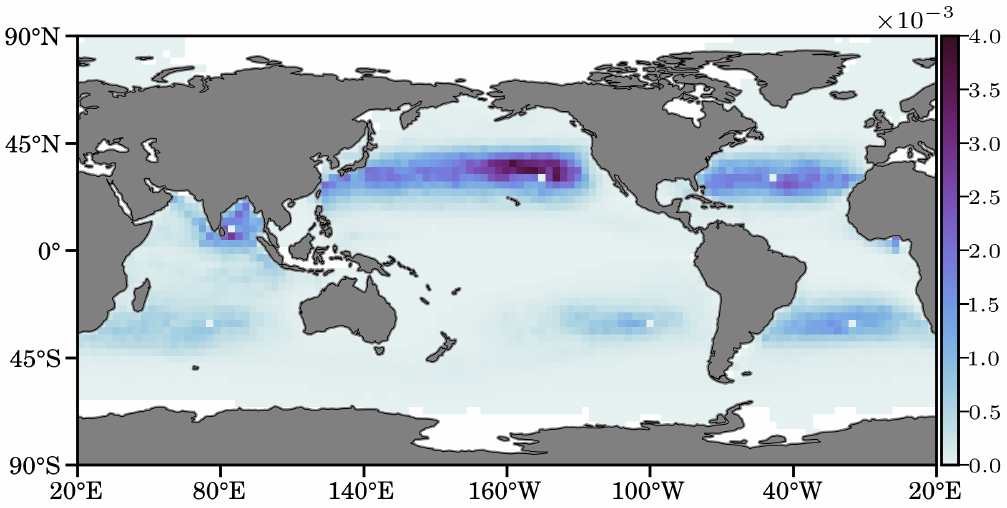}%
 \caption{Probability density of reactive debris paths that
 indicating where debris bottlenecks in their way into the garbage
 patches.}
 \label{fig:mu}%
\end{figure}

Additional insight is provided by the normalized distribution of
reactive trajectories $\hat{\boldsymbol\mu}^{AB}$, plotted in
Fig.\ \ref{fig:mu}, showing where reactive trajectories spend most
of the time while on their way from source $A$ to target $B$. Note
that the reactive trajectories tend to bottleneck over large regions
around the subtropical gyre patches except the Indian Ocean gyre
patch.  These regions measure the size of the patches.  The bottleneck
is particularly pretty intense in the North Pacific gyre.  The
reactive flows in the Indian Ocean patch are not seen to spend as
much time near the patch as near the other subtropical gyre patches.
This is consistent with it being a weak garbage patch. Additional
intense bottlenecks are observed to concentrate in the Bay of Bengal.

\begin{figure}[tb]
 \centering%
 \includegraphics[width=\columnwidth]{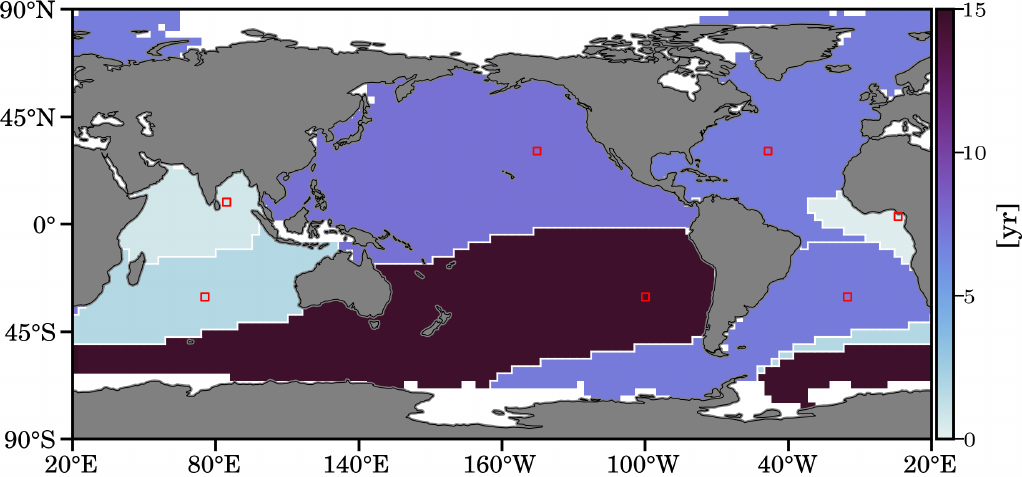}%
 \caption{Forward-committor-based dynamical geography revealing
 domains of influence for the garbage patches with the provinces
 colored according to residence time.}
 \label{fig:q}%
\end{figure}

Further insight into the domain of influence of each individual
garbage patch $g\in G$, and thus into the locations on the coast
where debris flows into them originate from, is offered by associating
to each state $i\in O$ the most likely patch $g$ (target) to hit
according to the probability in $i$ to forward-commit to $g$, viz.,
\begin{equation}
   q^+_i(g) = \Pr(\tau^+_g < \tau^+_{\omega} \mid X_0 = i).
  \label{eq:qb}
\end{equation}
This way every box of the partition gets assigned to a patch, forming
what we call a \emph{forward-committor-based dynamical geography},
which is shown in Fig.\ \ref{fig:q}.  Note the large influence exerted
by the subtropical patches on the global transport of marine debris,
particularly those in the subtropical gyres whose provinces span
the largest areas.  Similar influence of the subtropical patches
was inferred from spectral analysis \citep{Dellnitz-Junge-99,
Koltai-10} applied on simulated trajectories \citep{Froyland-etal-14}
and from direct evolutions using drifter trajectory
data.\citep{Maximenko-etal-12} The relatively large influence of
the Bay of Bengal patch inferred from the visual inspection of the
reactive currents into it is well framed by the geography.

The provinces of the geography in Fig.\ \ref{fig:q} are colored
according to the mean \emph{residence time}, defined as follows.
Let $Q\subset O$ be the box indices of a given province. The mean
time it takes a trajectory initialized in $i\in Q$ to move out of
$Q$ and thus hit the complement of $Q$,  $h^{Q}_i :=
\mathbb{E}(\tau^+_{O\cup\omega\setminus Q} \mid X_0 = i)$, is given
by the solution of the linear equation \cite{Norris-98, Dellnitz-etal-09,
Miron-etal-19-JPO} 
\begin{equation}
  \big(\Id^{|Q|\times|Q|} - P|_Q\big)\mathbf h^Q = \mathbf{1}^{|Q|\times
  1},
\end{equation}
where $\mathbf h^{Q} = \smash{(h_i^Q)_{i\in Q}}$.  By taking the
average of $\mathbf h^{Q}$ with respect to the stationary density
$\bpi\vert_Q$ we get the residence time in $Q$, i.e.,
\begin{equation}
  H^{Q} := \mathbb{E}(\tau^+_{O\cup\omega\setminus Q} \mid X_0 \in
  Q) = \frac{\mathbf h^{Q} \cdot \bpi\vert_Q}{\bpi\vert_Q \cdot
  \mathbf 1^{|Q|\times 1}}.
 \label{eq:hq}
\end{equation}
The longest residence time is 14.6 yrs, computed for the South
Pacific province, whereas the shortest residence times are 0.7 and
0.3 yrs for the Bay of Bengal and Gulf of Guinea regions, respectively.
The North Pacific Ocean, North Atlantic Ocean and South Atlantic
Ocean subtropical subtropical garbage patches all have comparable
residence times that range between 7--7.5 yrs while the Indian Ocean
garbage patch has a much shorter residence time of 1.8 years.

\subsection{Pollution paths out of subtropical garbage patches}

\begin{figure}[htp]
 \centering%
 \includegraphics[width=\columnwidth]{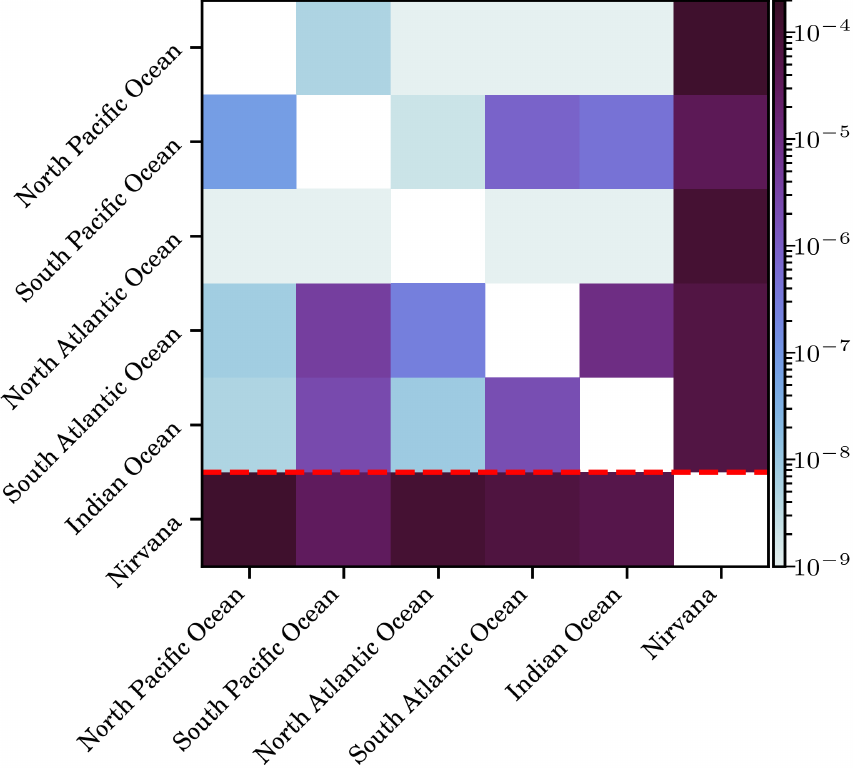}
 \caption{Reactive rates from each subtropical gyre patch, presented
 by row, into all other subtropical gyre patches and the nirvana
 state. The last row shows the rates from the nirvana state $\omega$
 into the subtropical gyre patches from the results presented on
 Fig.\ \ref{fig:f}.} \label{fig:kg}
\end{figure}

The interconnectivity of the subtropical garbage patches with respect
to the amount of debris particles that are exchanged between patches
is presented in Fig.\ \ref{fig:kg}. More precisely, we compute the
reactive flux from $A = y \in Y \subset G$ to $B = (Y \setminus y)
\cup \omega$, where $Y$ is the set of subtropical gyre patches.
Then, the proportions of total debris mass present in the ocean
that flow per time step out of $A$ and make their way towards
$B$ are $k^{AB}  =  1.4 \times 10^{-4}$, $3.7 \times 10^{-5}$, $1.1
\times 10^{-4}$, $6.2 \times 10^{-5}$, and $6.3 \times 10^{-5}$ for
$A$ chosen as the North Pacific, South Pacific, North Atlantic,
South Atlantic, and Indian Ocean patches, respectively.  We can
further decompose the transition rate from $A$ to $B$ into the sum
of arrival rates into each individual patch $b$ in $B$, $k^{AB} =
k^{B\leftarrow} = \sum_{b \in B} k^{b\leftarrow}$  as in
\eqref{eq:decomp_rate}.  For a fixed $A$, the arrival rates into
each $b\in B$ are shown in the rows of Fig.\ \ref{fig:kg}.  Consistently,
the ``emission'' from a garbage patch $y$ recirculates almost
completely through $\omega$ before reaching any other patch $Y
\setminus y$, hence the much higher rates in the column corresponding
to the nirvana state. In addition, relatively high reactive rates
between the subtropical garbage patches of the southern hemisphere
highlight an interconnection between the Indian Ocean, the South
Atlantic and the South Pacific patches.  Specifically, the Southern
Atlantic debris transit at high rate to the South Pacific and Indian
Oceans and, similarly, the Indian Ocean debris transit at high rate
to the South Pacific and Atlantic Ocean garbage patches. Finally,
the reactive rates from the North Atlantic gyre to any other
subtropical garbage patches are negligible, confirming again that
it has very little connection with other patches and debris that
manage to escape it most likely end up on land or in ice.

\begin{figure}[tb]
 \centering%
 \includegraphics[width=\columnwidth]{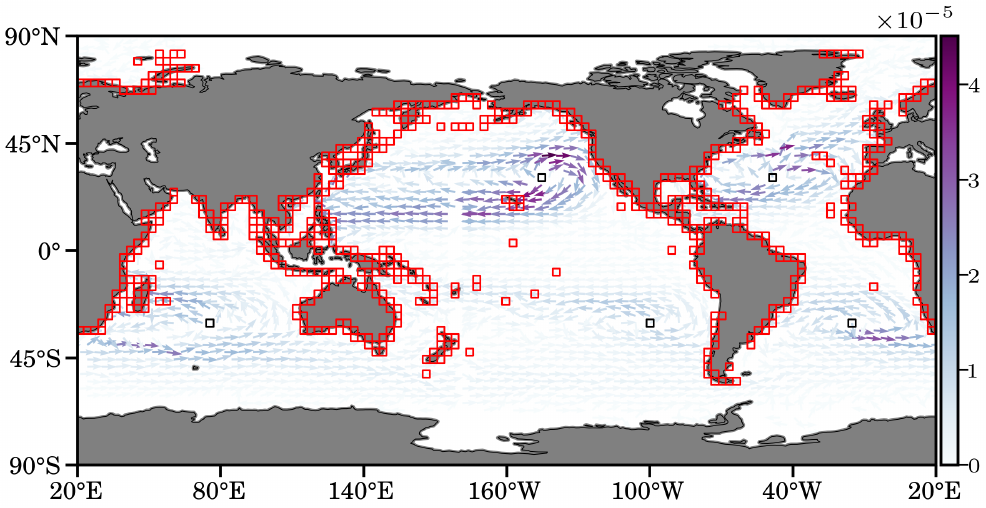}%
 \caption{Reactive currents from each subtropical gyre patch to the
 nirvana state $\omega$. The source set is indicated in black in
 each panel; the target sets are indicated in red.} 
 \label{fig:fomega}
\end{figure}

\begin{figure}[tb]
 \centering \includegraphics[width=\columnwidth]{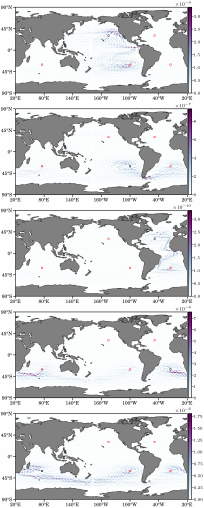}
 \caption{Reactive currents from each subtropical gyre patch to all
 other subtropical gyre patches. The source set is indicated in
 black in each panel; the target sets are indicated in red. Note
 the difference in the scales across the panels.} 
 \label{fig:f2}
\end{figure}

The last row of Fig.\ \ref{fig:kg} shows transition rates from
\eqref{eq:decomp_rate} corresponding to the currents into each
subtropical patches presented Fig.\ \ref{fig:f} with $A$ chosen as
the nirvana state $\omega$. As expected, the transition rates from
$\omega$ to the subtropical garbage patches are orders of magnitude
higher than the transition rates between patches. Bearing in mind
that those transition rates are very low, meaning that the transitions
are unlikely, associated reactive currents are depicted in
Figs.\ \ref{fig:fomega} and~\ref{fig:f2}.  These represent potential
pathways that marine debris might take out of the gyres, for instance,
in the event of unusually strong winds.

Figure~\ref{fig:fomega} presents the reactive currents from the
subtropical gyre patches to the nirvana state $\omega$, which
correspond to the last column of Fig.\ \ref{fig:kg}.  That is, we
set $A = Y$ (black squares) and $B = \omega$ (red squares are coastal
bins $i \in L$ where $P^{O \to \omega} > 0$). In general, debris
out of the northern hemisphere patches have a larger probability
of beaching than the southern hemisphere patches. In particular,
the reactive currents in the Indian Ocean follow the general path
of debris from the search area of the infamous Malaysia Airlines
flight MH370 to the locations of recovered debris on the coasts of
Mauritius, Madagascar, Mozambique, Tanzania, and South
Africa.\cite{Miron-etal-19-Chaos}

In turn, Fig.\ \ref{fig:f2} presents the reactive currents from a
subtropical gyre patch to the union of all other subtropical gyre
patches. To place the focus on debris trajectories that stay in the
ocean, we do not allow reactive passages via $\omega$. Thus we use
TPT for open domains by setting $A = y$ (black square) and $B =
Y\setminus y$ (red squares) in \eqref{eq:q+_open} and \eqref{eq:q-_open}.
The reactive currents out of the Indian Ocean patch are quite strong,
in agreement with reports \cite{vanderMheen-etal-19} on its weak
character.  However, these are somewhat weaker than those out of
the South Atlantic patch. Note that both the Indian Ocean patch and
the South Atlantic patch exchange debris with the South Pacific
Ocean patch, as shown in Fig.\ \ref{fig:kg}, through the Antarctic
Circumpolar Current. The currents that flow out of the North Pacific
patch are much weaker, yet not as weak as those coming out of the
North Atlantic patch. The strength of the currents out of the South
Pacific patch ranges in between the above.

To quantify the above qualitative conclusions from the inspection
of the transition channels, we computed the reactive rates from
each $y\in Y$ to $Y\setminus y$, telling us the amount of debris
probability mass that flows out of $y\in Y$ per time step and is
on its direct way to $Y\setminus y$, equal to the row sums of
Fig.\ \ref{fig:kg} excluding the portion that goes into nirvana. The
reactive rate \eqref{eq:kAB} gives $5.6 \times 10^{-9}$, $1.3 \times
10^{-6}$, $3.7 \times 10^{-10}$, $1.4 \times 10^{-5}$, and $4.5
\times 10^{-6}$ for the North Pacific, South Pacific, North Atlantic,
South Atlantic, and Indian Ocean patches, respectively, which confirm
our qualitative assessments above. We note that each of these rates
is at least one order of magnitude smaller than those reported at
the very beginning of this section, except that of the South Atlantic,
where it is merely a factor 5 weaker. This indicates that debris
leaving the South Atlantic is most frequently finding its way to
other patches.

It must be noted that the above reactive rates do not say anything
about retention. They tell us which patch ``emits'' the most
frequently such debris that finds its way to another patch. A low
rate does not need to mean that debris leaving $A$ comes back to
$A$ since the debris can hit $\omega$ too before hitting $B$, as
shown by the much higher reactive rates to the nirvana state in
Fig.\ \ref{fig:kg}. In other words, a low rate should not be taken
to mean the same as high attraction. Thus, the reactive rate computation
results just described do not contradict those from direct density
evolution in Fig.\ \ref{fig:1APk}, which had identified the North
Pacific patch as the most attracting of all.

\section{Summary and conclusions}\label{sec:conclusion}

We have presented a novel application of transition path theory
(TPT), here extended to open autonomous dynamical systems. The
problem chosen was that of pollution routes from possible coastline
sources (reactive states) into garbage patches in the global surface 
ocean (product states).

Undrogued drifter trajectories from NOAA Global Drifter Program
were used to derive a Markov chain on which TPT was applied, as a
model for the time-asymptotic dynamics of marine debris pollution.
Modeling the probability of trajectories to beach as a function of
the fraction of land filling each coastal box of the covering of
the world ocean domain resulted in an open system, which was closed
by sending the probability imbalance back into the chain according
to the capacity of coastal boxes to ``pollute the oceans'' as
measured by its share of global mismanaged plastic waste. Assuming
a constant pollution rate, our time-homogeneous model was shown to
be the statistical equivalent of a ``saturated'' (stationary)
pollution redistribution dynamics.

A high probability transition channel was identified connecting the
Great Pacific Garbage Patch with the coasts of Eastern Asia,
suggesting an important source of plastic pollution there.  The
weakness of the Indian Ocean gyre as a trap of plastic debris was
found consistent with transition paths not converging in the gyre.
While the North Pacific subtropical gyre was found to be most
attracting consistent with earlier assessments, the South Pacific
gyre stood out as the most enduring in the sense that the total
reactive rate out of that gyre into other gyres and the nirvana
state resulted the smallest of all.  The weakest of all the gyres
in terms of its capacity to trap and hold within plastic waste
resulted to be South Atlantic gyre. The gyres were found in general
weakly communicated.  Indeed, in the event of anomalously intense
winds a subtropical gyre is more likely to export garbage out toward
the coastlines than into another gyre.

Our results, including prospects for garbage patches yet to be
directly and/or robustly observed, namely, the Gulf of Guinea and 
the Bay of Bengal, have implications for activities such as ocean
cleanup as the revealed reactive pollution routes provide targets,
alternative to the great garbage patches themselves, to aim those
efforts.  Additional ocean applications of TPT are underway (e.g.,
using submerged float data and targeting meridional overturning
routes) and will be reported elsewhere.

\section*{Supplementary material}

The supplementary material contains versions of Figs.\ \ref{fig:1APk}
and~\ref{fig:f} assuming $\alpha = \smash{\frac{1}{2}}$ (Figs.\ S1
and~S4, respectively), $\smash{\frac{3}{4}}$ (Figs.\ S2 and~S5),
and $1$ (Figs.\ S3 and~S6) in \eqref{eq:PO}.

\begin{acknowledgments}
 We thank Mar\'ia J.\ Olascoaga for the benefit of many discussions
 on Lagrangian ocean dynamics. The data that support the findings
 of this study are openly available in the NOAA Global Drifter
 Program data set, available at http://\allowbreak
 www.aoml.noaa.gov/\allowbreak phod/\allowbreak dac/. The mismanaged
 plastic waste data are distributed from https://\allowbreak
 ourworldindata.org/. The numerical code to reproduce the findings
 of this study is openly available at https://\allowbreak
 github.com/\allowbreak philippemiron/pygtm.  Support for this work
 was provided by NSF grant OCE1851097 (PM, FJBV), Deutsche
 Forschungsgemeinschaft (DFG) through grant CRC 1114 ``Scaling
 Cascades in Complex Systems'', Project Number 235221301, Project
 A01 (PK) and Germany's Excellence Strategy -- The Berlin Mathematics
 Research Center MATH+ (EXC-2046/1, project ID: 390685689) (LH).
\end{acknowledgments}

%

\clearpage

\renewcommand\thefigure{S\arabic{figure}}    
\setcounter{figure}{0}  

\begin{figure*}[h!]
 \centering%
 \includegraphics[width=.6\textwidth]{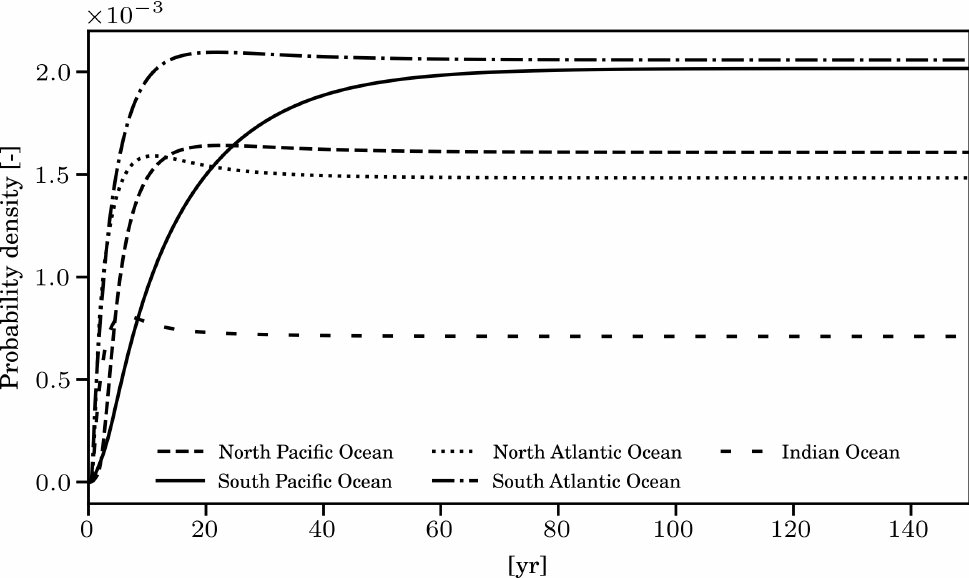}%
 \caption{As in Fig.\ 4, but assuming $\alpha = \frac{1}{2}$ in
 (28).}
\end{figure*}

\begin{figure*}[h!]
 \centering%
 \includegraphics[width=.6\textwidth]{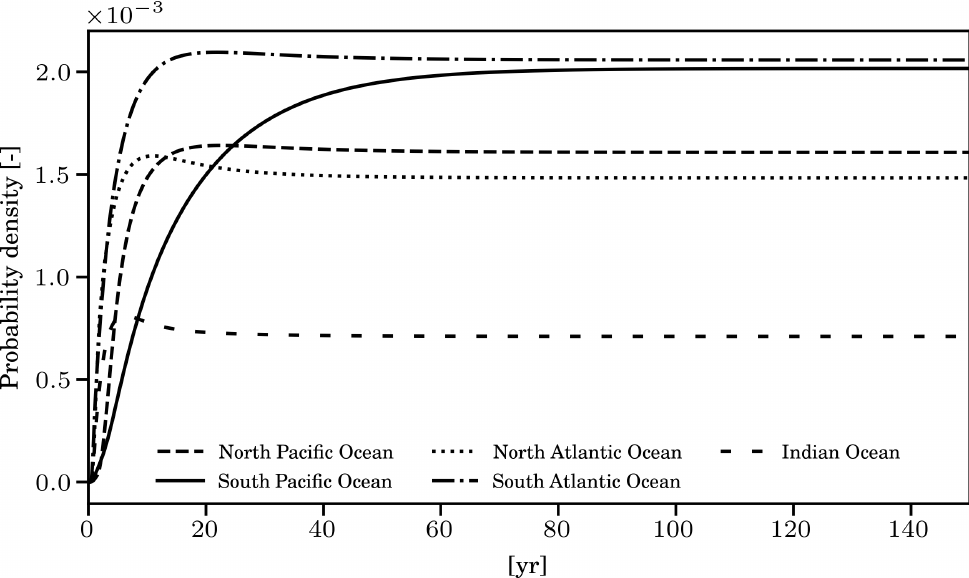}%
 \caption{As in Fig.\ 4, but assuming $\alpha = \frac{3}{4}$ in
 (28).}
\end{figure*}

\begin{figure*}[h!]
 \centering%
 \includegraphics[width=.6\textwidth]{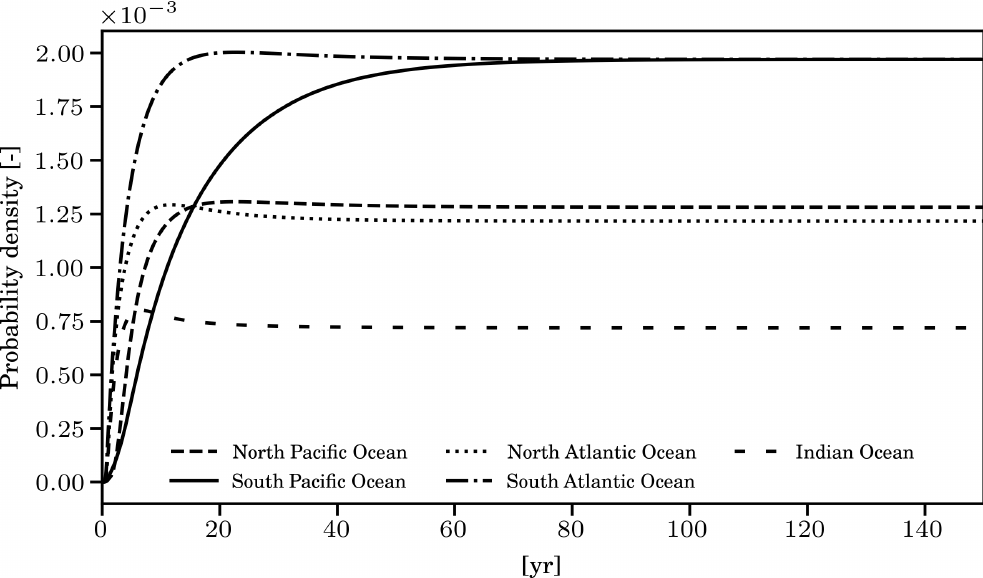}%
 \caption{As in Fig.\ 4, but assuming $\alpha = 1$ in (28).}
\end{figure*}

\clearpage

\begin{figure*}[h!]
 \centering%
 \includegraphics[width=.75\textwidth]{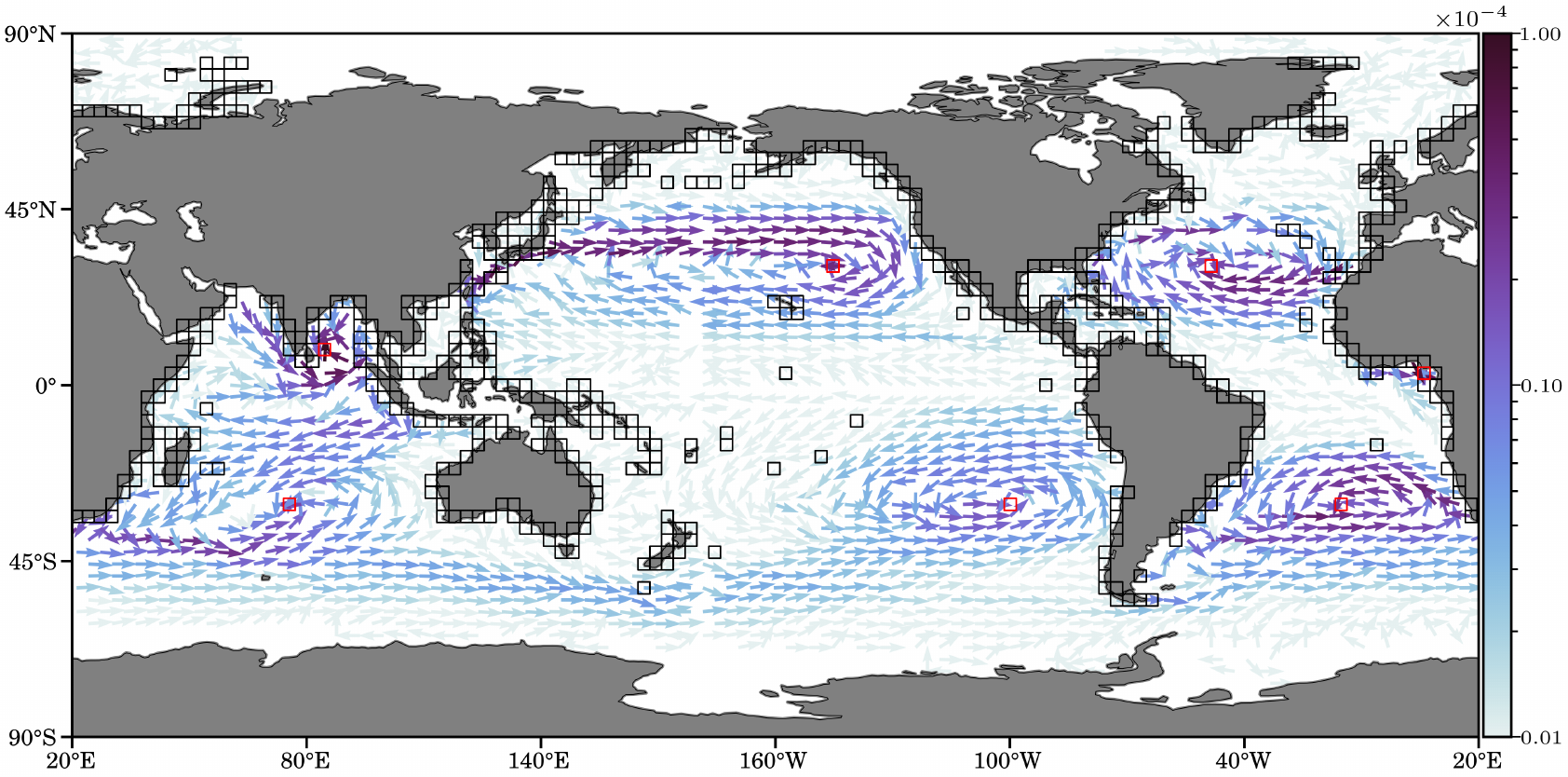}%
 \caption{As in Fig.\ 5, but assuming $\alpha = \frac{1}{2}$ in
 (28).}
\end{figure*}

\begin{figure*}[h!]
 \centering%
 \includegraphics[width=.75\textwidth]{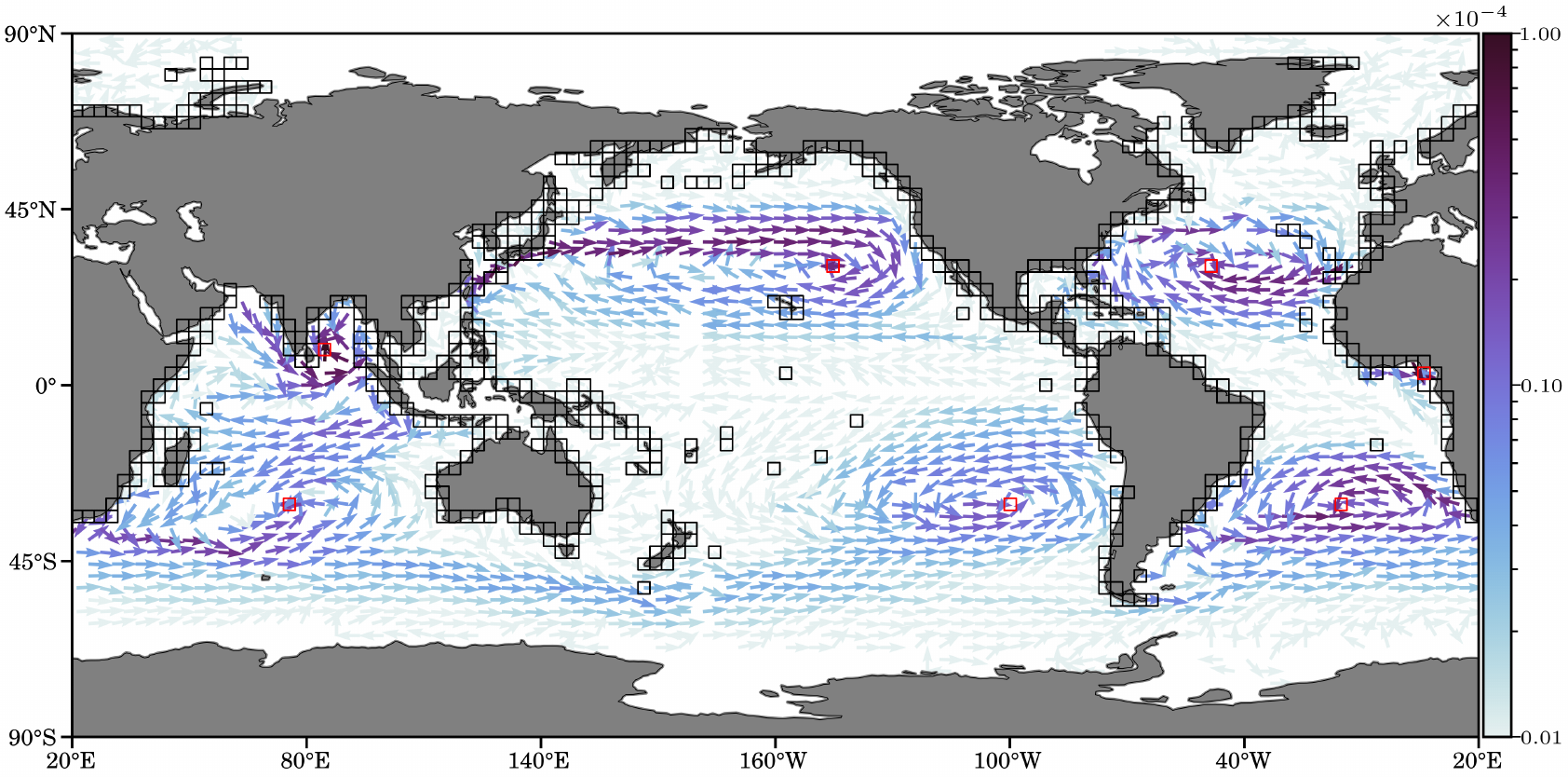}%
 \caption{As in Fig.\ 5, but assuming $\alpha = \frac{3}{4}$ in
 (28).}
\end{figure*}

\begin{figure*}[h!]
 \centering%
 \includegraphics[width=.75\textwidth]{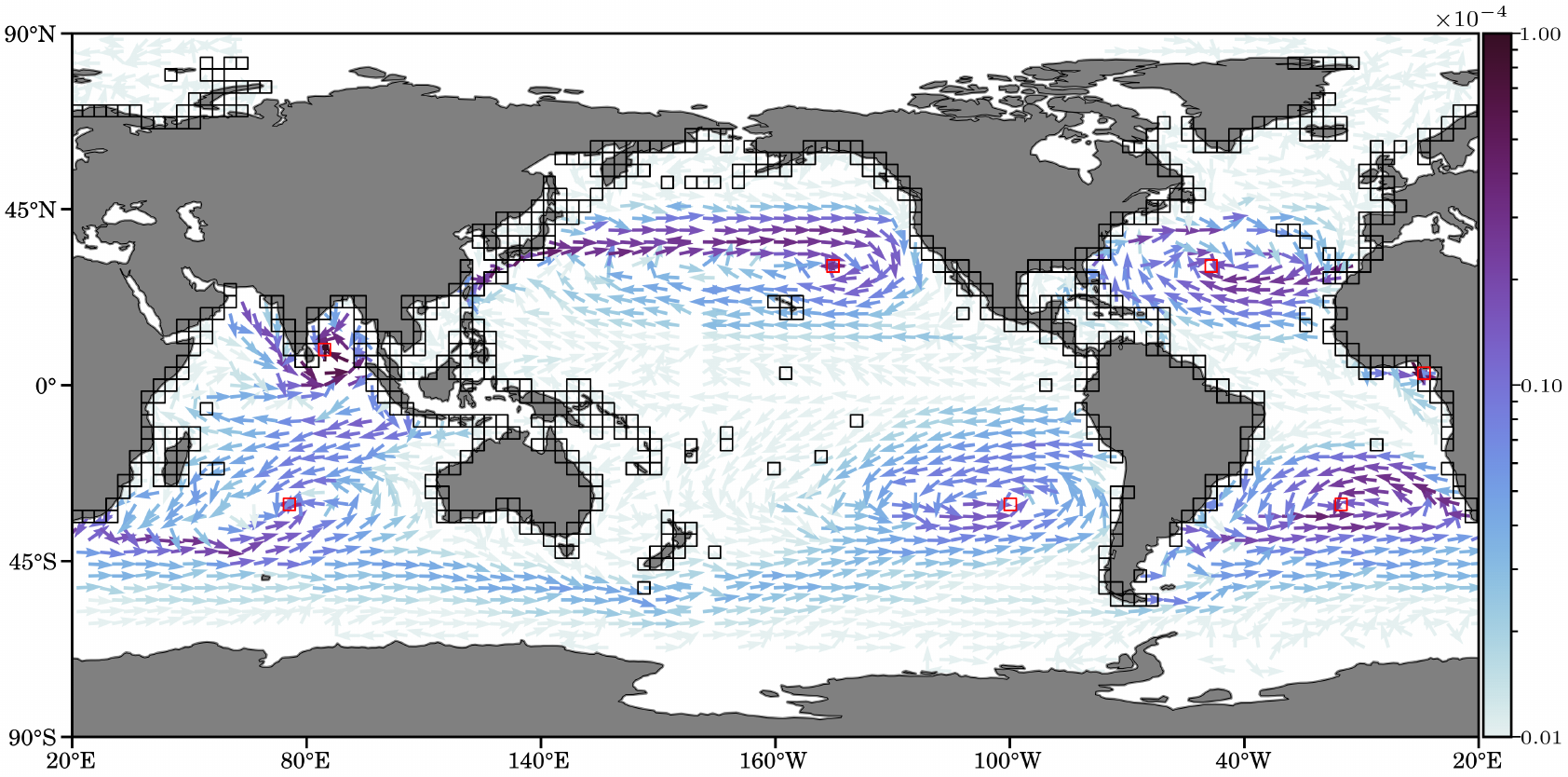}%
 \caption{As in Fig.\ 5, but assuming $\alpha = 1$ in (28).}
\end{figure*}

\end{document}